\documentclass[pra,twocolumn,amsmath,amssymb,superscriptaddress]{revtex4-1}
\usepackage{epsfig,amsmath}
\usepackage{subfigure}
\usepackage{graphicx}% Include figure files
\usepackage{dcolumn}% Align table columns on decimal point
\usepackage{stmaryrd}
\usepackage{mathrsfs}
\usepackage{pifont}
\usepackage{amsthm}
\usepackage{amssymb}
\usepackage{bm}
\usepackage{latexsym}
\usepackage[colorlinks=true,linkcolor=blue,citecolor=blue]{hyperref}
\usepackage{color}
\usepackage{epstopdf}
\usepackage{booktabs}

\begin{document}

\title{Magnon blockade in magnon-qubit systems}

\author{Zhu-yao Jin}
\affiliation{School of Physics, Zhejiang University, Hangzhou 310027, Zhejiang, China}

\author{Jun Jing}
\email{Email address: jingjun@zju.edu.cn}
\affiliation{School of Physics, Zhejiang University, Hangzhou 310027, Zhejiang, China}

\date{\today}

\begin{abstract}
A hybrid system established by the direct interaction between a magnon mode and a superconducting transmon qubit is used to realize a high-degree blockade for magnon. It is a fundamental way toward quantum manipulation at the level of a single magnon and preparation of single magnon sources. Through weakly driving the magnon and probing the qubit, our magnon-blockade proposal can be optimized when the transversal coupling strength between the magnon and qubit is equivalent to the detuning of the qubit and the probing field or that of the magnon and the driving field. Under this condition, the equal-time second-order correlation function $g^{(2)}(0)$ can be analytically minimized when the probing intensity is about three times the driving intensity. Moreover, the magnon blockade could be further enhanced by proper driving intensity and system decay rate, whose magnitudes outrange the current systems of cavity QED and cavity optomechanics. In particular, the correlation function achieves $g^{(2)}(0)\sim10^{-7}$, about two orders lower than that for the photon blockade in cavity optomechanics. Also, we discuss the effects on $g^{(2)}(0)$ from thermal noise and the extra longitudinal interaction between the magnon and qubit. Our optimized conditions for blockade are found to persist in these nonideal situations.
\end{abstract}

\maketitle

\section{Introduction}

Strong coupling among distinct quantum systems allows for efficient quantum state engineering, which plays a crucial role in quantum information processing~\cite{Wallquist2009Hybrid,Kimble2008internet} and quantum networks~\cite{Kimble2008internet}. Beyond the conventional paradigm of the strong coupling to the optical modes via large electric dipole moments of the system, it was found that a strong light-matter interaction can also be established via magnetic dipole moments~\cite{Imamo2009Cavity}. In the past decade, a great deal of attention has been focused on various ferromagnetic systems with high spin densities, including ultracold atomic clouds~\cite{Verd2009Strong}, molecule nanomagnets in a cavity~\cite{Eddins2014collective}, nitrogen vacancy centers in diamond~\cite{Kubo2010Strong,Ams2011Cavity,Marcos2010coupling,Ranjan2013probing}, and ion-doped crystals~\cite{Schuster2010high,Probst2013Anisotropic,Tkal2014strong}. Hybrid magnon systems of yttrium iron garnet (YIG) spheres~\cite{Tabuchi2014Hybridizing,Huebl2013High,Zhang2014Strongly,Goryachev2014High,
Bai2015Spin,Soykal2010Strong,Bourhill2016Ultrahigh,Zare2015Magnetic,Viola2016Coupled,
Zhang2016Opto,Osada2016Cavity,Haigh2016Triple} demonstrate distinguishable advantages by virtue of a high order of magnitude of spin density ($\rho_s=4.22\times10^{27}$ m$^{-3}$), an extremely low damping rate of magnons (the quanta of collective spin excitations), and the nonlinear amplification and control of magnons via the interaction between spin excitations. Recent theories~\cite{Liu2019Magnon,Xie2020Quantum} and experiments~\cite{Yutaka2015Coherent,Dany2017Resolving,Wolski2020Dissipation,Dany2020Entanglement,Kounalakis2022Analog} suggest that the magnon-qubit system could be a promising information carrier since indirect (mediated by a microwave cavity)~\cite{Liu2019Magnon,Xie2020Quantum,Yutaka2015Coherent,Dany2017Resolving,Wolski2020Dissipation,Dany2020Entanglement} and direct~\cite{Kounalakis2022Analog} interactions between a YIG sphere and a superconducting qubit have been established.

Many outstanding quantum properties in the hybrid systems of magnons, such as polariton bistability~\cite{wang2018bistability} and distant spin currents~\cite{Bai2017Cavity}, have been detected on a mesoscopic scale. Owing to these advances, growing interest has arisen in understanding the properties on the microscopic scale, including cooling magnons by a monochromatic laser source~\cite{Sharma2018Optical}, storing information in the magnon dark modes~\cite{Zhang2015magnon}, entangling magnon-magnon system in a cavity~\cite{Yuan2020enhancement,Azimi2021Magnon} or through a microwave cavity array~\cite{Ren2022Long}, generating the magnon-photon-phonon entanglement~\cite{Li2018Magnon}, preparing Bell state between magnon and photon~\cite{Yuan2020Steady}, and creating a superposed state of a single magnon and vacuum~\cite{Xu2022Quantum}. Many of these works are based on the assumption that the magnon can be prepared at or close to the Kittel mode, i.e., the ground state, and then can be operated at the level of a single magnon. As a pure quantum effect, blockade is another fundamental method to push the magnon system close to the ground state (yet it is not the target state) and may find applications in designing single magnon emitters. In general, a certain blockade is characterized by a nonvanishing single-excitation population and greatly suppressed multiple-excitation populations. Such a state is of great interest in the quantum realm of magnon systems and yet remains an outstanding challenge in practice. A group of well-known and important examples of quantum blockade can be found in optical systems~\cite{Zheng2011Cavity,Huang2013photon,Liao2010correlated,Ghosh2019Dynamical,Lu2015Squeezed,Peyronel2012quantum,
Reinhard2012strongly,Reinhard2012strongly,Liew2010Single,Bamba2011Origin,Snijders2018Observation,Vaneph2018Observation,
Hartmann2007strong,Chang2014Quantum}. Photon blockade prevents a second photon from entering the cavity mode~\cite{Imamo1997strongly,Hartmann2007strong,Chang2014Quantum,Tang2015Quantum}, leading to antibunching in photon correlation measurements. Similarly, single-magnon manipulation could be supported by the magnon blockade, which occurs under the anharmonicity in the energy spectrum.

In this work, we create and enhance the magnon blockade in a hybrid system, where a YIG sphere couples directly to a superconducting transmon qubit through the magnetic stray field~\cite{Kounalakis2022Analog}. This system features the transversal (exchange) and longitudinal interactions between the magnon and qubit, which are conveniently tunable by both the external flux through the superconducting quantum interference device (SQUID) loop and the magnon-qubit distance. A general goal in quantum blockade is to minimize the equal-time second-order correlation function $g^{(2)}(0)$, which is commonly used to characterize and distinguish the bunching and antibunching of the interested quanta. The expectation value of $g^{(2)}(0)$ is evaluated by the steady state of the magnon, which can be obtained by the master equation for the composite system~\cite{Carmichael1999statistical}. In the steady-state regime, the nonclassical antibunching effect can be observed when $g^{(2)}(0)<1$. It is a pure quantum scenario where the magnons have the tendency to be detected further away in space (time) than those of a classical state. To amplify the blockade effect, a probing field and a driving field are applied to the qubit and magnon, respectively. Our magnon-blockade proposal is verified and optimized under the particular conditions that the magnon-qubit transversal coupling strength and the detuning of the qubit (magnon) and the probing (driving) field are the same in magnitude and the intensity of the probing field is three times the driving field. The better blockade in the magnon system relies on a comparatively much stronger interaction between the magnon and qubit and a much lower damping rate of the magnon than their counterparts for photon blockade. The function $g^{(2)}$ in magnon blockade is found to be as low as approximately $10^{-7}$, about two orders lower than the minimum value in photon blockade~\cite{Wang2020Photon}. We find that the optimal conditions for the high-degree magnon blockade survive under the influences of thermal noise and the extra longitudinal interaction between the magnon and qubit.

The rest of this work is structured as follows. In Sec.~\ref{modelandHam} we introduce the hybrid magnon-qubit system under external driving and probing after a brief review of the physics underlying quantum blockade. In Sec.~\ref{OptimalDrivingStrength}, we investigate the optimal conditions for magnon blockade within the experimentally relevant regimes of Rabi frequencies of driving and probing fields and coupling strength between the magnon and qubit. An optimized driving intensity to minimize $g^{(2)}(0)$ of the steady state can be identified by choosing the proper transversal coupling strength and system decay rate. In Sec.~\ref{discussion} we discuss the persistence of the optimal conditions under two nonideal situations. We summarize the work in Sec.~\ref{Conclusion}. In the Appendix we provide details of the analytical model for blockade and the population dynamics of the magnon mode.

\section{model and Hamiltonian}\label{modelandHam}

\begin{figure}[htbp]
\centering
\includegraphics[width=0.95\linewidth]{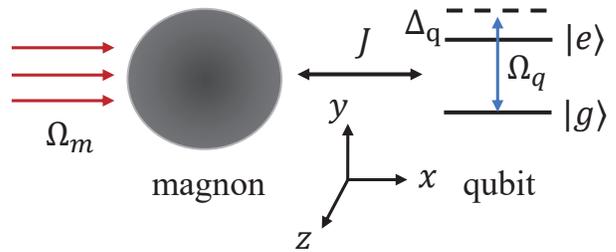}
\caption{Sketch of the magnon-qubit system, in which the ferromagnetic YIG sphere is directly coupled to the superconducting transmon qubit. The magnon mode and the qubit are under the driving field with Rabi frequency $\Omega_m$ and the probing field with Rabi frequency $\Omega_q$, respectively.}\label{model}
\end{figure}

A typical photon-blockade effect in cavity quantum electrodynamics (QED) relies on the strong energy-spectrum anharmonicity. Taking the classic Jaynes-Cummings (JC) model~\cite{Yutaka2015Coherent,Jaynes1963comparison} as an example, the strong coupling between the cavity and the two-level system induces a large vacuum Rabi splitting~\cite{Imamo1997strongly,Hartmann2007strong,Chang2014Quantum}, which gives rise to anharmonic ladders of the dressed states described by eigenvectors $|n,\pm\rangle$ with eigenvalues $E_{n,\pm}=n\omega\pm\sqrt{n}J$. Here $\omega$ and $J$ are the resonant frequency of both the photon and qubit and their coupling strength, respectively, $n$ denotes the excitation number, and $+$ ($-$) represents the higher (lower) branch. In this nonlinear energy spectrum, the transitions between $|0,g\rangle$ and $|1,\pm\rangle$ are allowed by the resonant driving, while the transitions $|1,\pm\rangle\rightarrow|2,\pm\rangle$ are suppressed by the energy mismatch, i.e., $2(\omega\pm J)\neq2\omega\pm\sqrt{2}J$.

Given the similarity in the anharmonic spectrum, our study is conducted on a hybrid system, in which the YIG sphere (magnon) is directly coupled to a superconducting transmon qubit~\cite{Kounalakis2022Analog} along the $x$ axis (see Fig.~\ref{model}). The original Hamiltonian reads ($\hbar\equiv1$)
\begin{equation}\label{Ham_initial}
H=\omega_q\sigma_+\sigma_-+\omega_mm^\dagger m+J(m\sigma_++m^\dagger\sigma_-),
\end{equation}
where $\sigma_+$ ($\sigma_-$) and $m^\dagger$ ($m$) are the creation (annihilation) operators of the transmon qubit and magnon mode with frequencies $\omega_q$ and $\omega_m$, respectively, and $J$ is the transversal (exchange) coupling strength between them. The qubit is formed by a SQUID loop and a capacitor in parallel. The SQUID loop is interrupted by two Josephson junctions. In the experiment of Ref.~\cite{Kounalakis2022Analog}, the flux through the SQUID loop consists of two parts: (i) the external flux $\Phi_b$ applied via control lines carrying direct and alternating currents and (ii) the flux $\Phi(\Delta\mu)$ induced by the magnetic fluctuations $\Delta\mu$. The latter establishes a direct magnon-qubit interaction, including the transversal and longitudinal parts, both of which are tunable by the external flux $\Phi_b$ and the magnon-qubit distance $d$. The transversal interaction, akin to the exchange interaction in the JC model~\cite{Yutaka2015Coherent,Jaynes1963comparison}, is a basic element of the magnon blockade. The longitudinal interaction, which is analogous to the radiation pressure in optomechanics~\cite{Shevchuk2017Strong,Rodrigues2019coupling,Kounalakis2020Flux}, can be omitted when fixing the externally applied flux on the qubit as $\Phi_b/\Phi_0\approx0.5$ ($\Phi_0$ is the magnetic flux quantum). Its effect on magnon blockade will be discussed in Sec.~\ref{secLongitudinal}. The coupling strength $J/2\pi$ for the flux on the qubit $\Phi_b/\Phi_0\approx0.5$ can be improved as much as about $40$ MHz~\cite{Kounalakis2022Analog} by modulating the magnon position. The magnon frequency $\omega_m/2\pi$ is about $1-10$ GHz and its decay rate $\kappa_m=\omega_m\alpha_G$ is about $0.1-1$ MHz in terms of the Gilbert damping constant $\alpha_G$. The qubit frequency $\omega_q/2\pi$ is widely tunable from $1$ to $10$ GHz by modulating the external flux and the SQUID asymmetry $a$~\cite{Kounalakis2022Analog}, which is about $1.5$ GHz for $\Phi_b/\Phi_0\approx0.5$ and $a=0.1$.

Jaynes-Cummings-like interaction is not sufficient to achieve a high degree of blockade. To enhance and manipulate the magnon blockade, we apply a probing field and a driving field on the qubit and magnon, respectively, as shown in Fig.~\ref{model}. The two fields are assumed to have the same frequency $\omega$, which is of the order of gigahertz~\cite{wang2018bistability}. Then the total Hamiltonian reads
\begin{equation}\label{Ham_tot}
\begin{aligned}
H_{\rm tot}&=\omega_q\sigma_+\sigma_-+\omega_mm^\dagger m+J(m\sigma_++m^\dagger\sigma_-)\\
&+\Omega_m(m^\dagger e^{-i\omega t}+m e^{i\omega t})+ \Omega_q(\sigma_+e^{-i\omega t}+\sigma_-e^{i\omega t}),
\end{aligned}
\end{equation}
where $\Omega_q=k\sqrt{P_d}$ with $k=103$ MHz/mW$^{1/2}$~\cite{wang2019simulation} and the field power $P_d$ is up to $350$ mW~\cite{wang2018bistability}, and $\Omega_m=\sqrt{2S}\Omega_s$, with $S$ the total spin number of the macrospin and $\Omega_s$ the coupling strength of the driving field with the macrospin~\cite{wang2016magnon}. The time dependence of the Hamiltonian can be removed in the rotating frame with respect to $H'=\omega(m^\dagger m+\sigma_+\sigma_-)$. Then the effective Hamiltonian of this hybrid system becomes
\begin{equation}\label{Ham_eff}
\begin{aligned}
H_{\rm eff}&=(\Delta_+-\Delta_-)\sigma_+\sigma_-+(\Delta_++\Delta_-)m^\dagger m\\
&+J(m\sigma_++m^\dagger\sigma_-)+\Omega_m(m^\dagger+m)+\Omega_q(\sigma_++\sigma_-),
\end{aligned}
\end{equation}
where $\Delta_{\pm}\equiv(\Delta_m\pm\Delta_q)/2$ and $\Delta_q\equiv\omega_q-\omega$ ($\Delta_m\equiv\omega_m-\omega$) is the detuning of the qubit (magnon) and the probing (driving) field.

\section{Magnon blockade and optimal conditions}\label{OptimalDrivingStrength}

The density matrix $\rho_m$ of the magnon mode is obtained by partially tracing the full density matrix $\rho$ over the degrees of freedom of the qubit, $\rho_m={\rm Tr}_q[\rho]$. Using a master-equation approach to take decoherence into account~\cite{Carmichael1999statistical}, we are able to study both the dynamics and the steady state of the full density matrix. The master equation reads,
\begin{equation}\label{masterequation}
\frac{\partial}{\partial t}\rho=-i[H_{\rm eff},\rho]+\frac{\kappa_m}{2}\mathcal{L}_m[\rho]+\frac{\kappa_q}{2}\mathcal{L}_{\sigma_-}[\rho],
\end{equation}
where the dissipator is defined as the Lindblad superoperators $\mathcal{L}_o[\rho]=2o\rho o^\dagger-o^\dagger o\rho-\rho o^\dagger o$~\cite{Scully1997quantum}, with $o=m, \sigma_-$ indicating the decay channels for the magnon mode and the qubit, respectively. The decay rates of the magnon and qubit are $\kappa_m$ and $\kappa_q$, respectively, which are set as  $\kappa_m=\kappa_q=\kappa$ for simplicity. The degree of the magnon blockade is characterized by the equal-time second-order correlation functions~\cite{Carmichael1999statistical,Eleuch2008Photon}
\begin{equation}\label{g2define}
g^{(2)}(0)=\frac{\langle m^\dagger m^\dagger mm\rangle}{\langle m^\dagger m\rangle^2},
\end{equation}
where the expectation value is calculated by $\rho_m$ after solving Eq.~(\ref{masterequation}). The steady-state correlation function $g^{(2)}(0)=1$ indicates the Poissonian distribution of the magnons. $g^{(2)}(0)>1$ indicates the super-Poissonian distribution corresponding to the magnon bunching effect and $0<g^{(2)}(0)<1$ indicates the sub-Poissonian distribution as a result of the magnon antibunching effect. The asymptotic limit $g^{(2)}(0)\rightarrow0$ describes the magnon blockade.

\begin{figure}[htbp]
\centering
\includegraphics[width=0.95\linewidth]{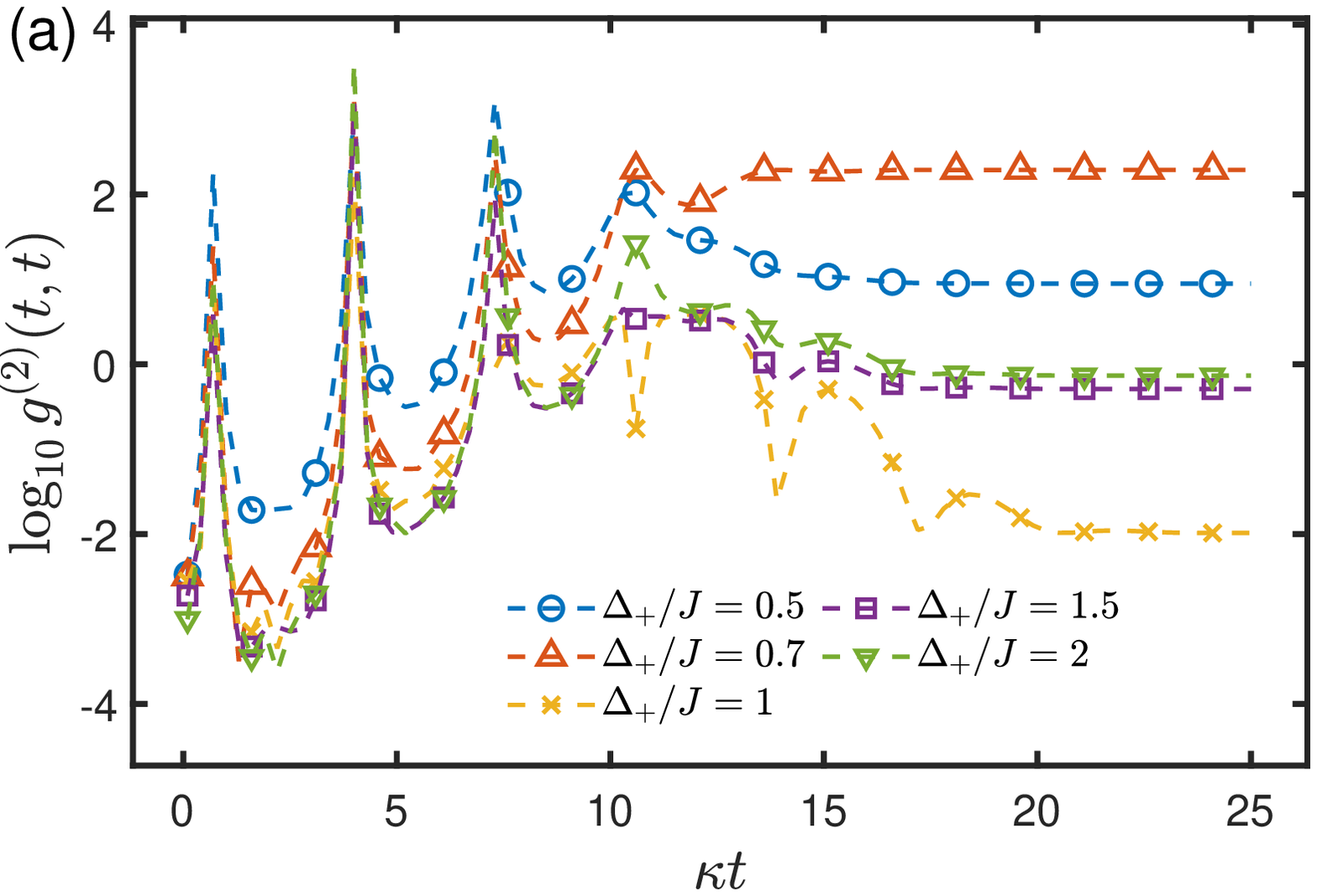}
\includegraphics[width=0.95\linewidth]{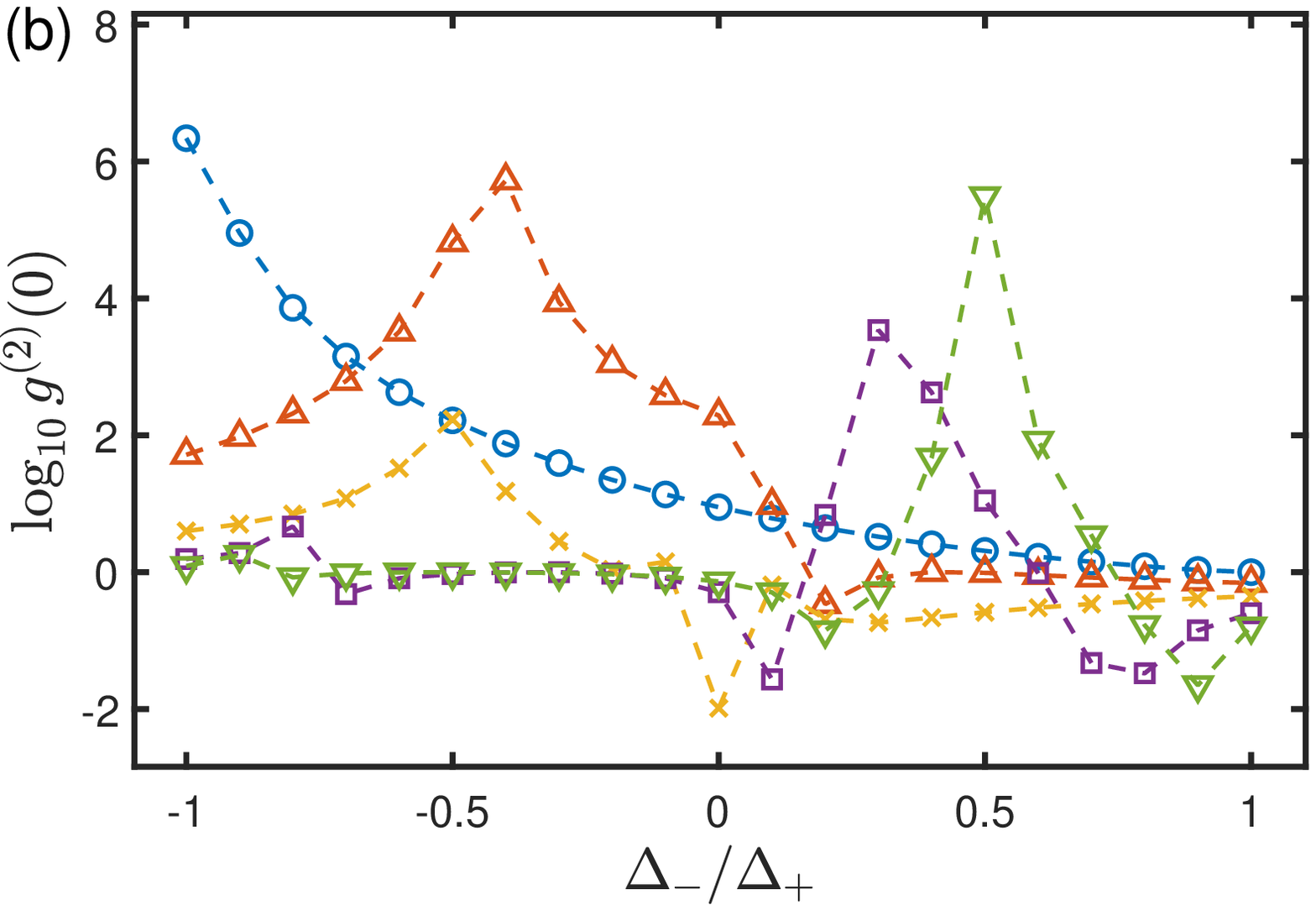}
\caption{(a) Time evolution of the logarithmic function of the second-order correlation function $g^{(2)}(t,t)\equiv\langle m^\dagger(t) m^\dagger(t) m(t)m(t)\rangle/\langle m^\dagger(t)m(t)\rangle^2$ under $\Delta_-=0$ with various $\Delta_+$. (b) Plot of $\log_{10}g^{(2)}(0)$ of the steady state as a function of the ratio $\Delta_-/\Delta_+$. The parameters are $J/2\pi=20$ MHz, $\Omega_m/2\pi=0.1$ MHz ($0.037$ $\mu$W), $\Omega_q=\Omega_m$, and $\kappa/2\pi=1$ MHz. }\label{dynamical}
\end{figure}

Due to Eq.~(\ref{Ham_eff}), the logarithmic function of the equal-time second-order correlation function $g^{(2)}$ relies on the average $\Delta_+$ of the two detunings $\Delta_m$ and $\Delta_q$ and their distance $\Delta_-$. To identify the optimizable parameters for the antibunching effect, we first demonstrate the behaviors of the function $g^{(2)}$ with respect to the temporal evolution and the steady state in Figs.~\ref{dynamical}(a) and \ref{dynamical}(b), respectively. In Fig.~\ref{dynamical}(a) for $\Delta_m=\Delta_q$, all of the functions $g^{(2)}$ approach their asymptotic values when $\kappa t\ge20$, i.e., $g^{(2)}(t,t)\rightarrow g^{(2)}(0)$ in the steady-state regime. The antibunching effect emerges when $\Delta_+/J$ is close to unit. In particular, the function $g^{(2)}$ finds its minimum value as $g^{(2)}(t,t)\rightarrow g^{(2)}(0)\sim10^{-2}$ when $\Delta_+/J=1$, which motivates us to explore the optimal magnitude of the off-resonance $\Delta_-$ under a fixed $\Delta_+$. In Fig.~\ref{dynamical}(b) for the steady value, it is found that the function $g^{(2)}$ takes the minimum value when $\Delta_-/\Delta_+=0$. A nonzero $\Delta_-$ induces a negative effect on blockade. Then we set $\Delta_-=0$, i.e., $\Delta_q=\Delta_m$ or $\omega_q=\omega_m$, in the following exploration.

\begin{figure}[htbp]
\centering
\includegraphics[width=0.95\linewidth]{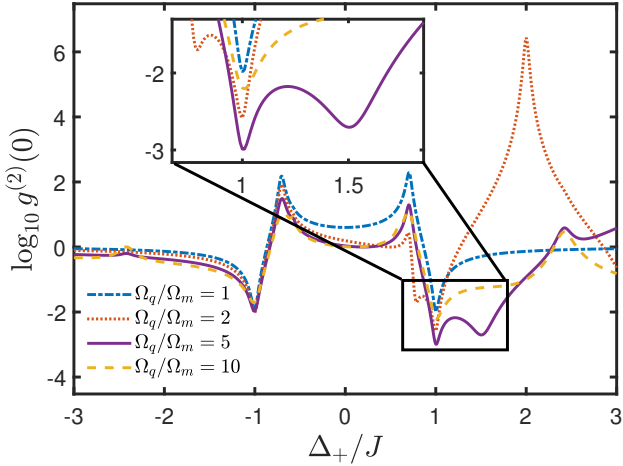}
\caption{Steady-state $\log_{10}g^{(2)}(0)$ versus the ratio of the average detuning to the transversal coupling strength $\Delta_+/J$. The parameters are $J/2\pi=20$ MHz, $\Omega_m/2\pi=0.1$ MHz, and $\kappa/2\pi=1$ MHz.}\label{conditionDelta}
\end{figure}

To confirm the conditions under which the antibunching effect occurs, we show in Fig.~\ref{conditionDelta} the dependence of $\log_{10}g^{(2)}(0)$ on the ratio of the average detuning and the magnon-qubit transversal coupling strength $\Delta_+/J$ for various ratios of field intensities (Rabi frequencies) $\Omega_q$ and $\Omega_m$. The intensities of both probing and driving fields are experimentally tunable through modulating the field power~\cite{wang2018bistability}. In Fig.~\ref{conditionDelta} the quantum-classical boundary $g^{(2)}(0)=1$ separates the bunching regime [$g^{(2)}(0)>1$] and the antibunching regime [$g^{(2)}(0)<1$]~\cite{Rabl2011photon}. For all ratios of $\Omega_q/\Omega_m$, a strong bunching effect occurs around $\Delta_+/J\approx\pm0.7$, where $g^{(2)}(0)\sim10^{2}$, and a strong antibunching effect occurs around $\Delta_+/J=\pm1$. When $\Delta_+/J=-1$, one can find that the expectation value of the function $g^{(2)}$ is not sensitive to $\Omega_q/\Omega_m$. In contrast, a higher degree of magnon blockade favors the condition $\Delta_+/J=1$ and it is able to be modulated by the ratio of the probing-field and driving-field intensities. In particular, we have $g^{(2)}(0)=10^{-1.99}$, $g^{(2)}(0)=10^{-2.58}$, $g^{(2)}(0)=10^{-2.99}$, and $g^{(2)}(0)=10^{-2.21}$ for $\Omega_q/\Omega_m=1$, $\Omega_q/\Omega_m=2$, $\Omega_q/\Omega_m=5$, and $\Omega_q/\Omega_m=10$, respectively. We are therefore guided to investigate an optimized condition about $\Omega_q/\Omega_m$ to further enhance the degree of the magnon blockade.

\begin{figure}[htbp]
\centering
\includegraphics[width=0.95\linewidth]{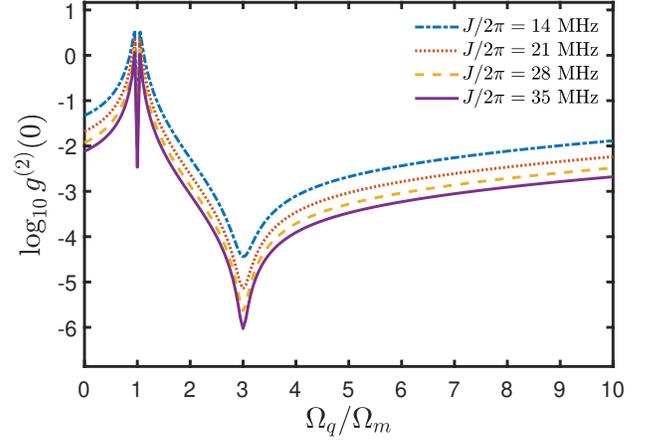}
\caption{Steady-state $\log_{10}g^{(2)}{(0)}$ versus $\Omega_q/\Omega_m$ for $\Delta_+/J=1$, $\Omega_m/2\pi=0.1$ MHz, and $\kappa/2\pi=1$ MHz.}\label{conditionDrive}
\end{figure}

In Fig.~\ref{conditionDrive}, with various transversal coupling strengths $J$, we show the dependence of $\log_{10}g^{(2)}(0)$ on the ratio of probing intensity and driving intensity for $\Delta_+=J$. It is interesting to find that for all $J$'s, the function $g^{(2)}$ might take local minimum values when $\Omega_q/\Omega_m=1$ and $\Omega_q/\Omega_m=3$. The latter can be identified as a global minimum point, around which $g^{(2)}(0)$ decreases monotonically with increasing $J$. The transversal interaction strength is upper bounded by state-of-the-art experiments (about $J/2\pi\sim40$ MHz~\cite{Kounalakis2022Analog}). Nevertheless, the degree of magnon blockade is enhanced from $g^{(2)}(0)=10^{-4.44}$ for $J/2\pi=14$ MHz to $g^{(2)}(0)=10^{-6.03}$ for $J/2\pi=35$ MHz.

The ratio of the Rabi frequency of the probing field $\Omega_q$ and that of the driving field $\Omega_m$ in capturing the minimum function $g^{(2)}$ could be self-consistently estimated by an analytical model~\cite{Kar2013Single} as follows. In the weak-driving limit in comply with the ignorance of the quantum jump during a short timescale, the master equation~(\ref{masterequation}) can be simulated with the non-Hermitian Hamiltonian $H_{\rm non}=H_{\rm eff}-i\kappa(\sigma_+\sigma_-+m^\dagger m)/2$. Consequently, the composite system could be truncated to a subspace with a few excitations and the system evolution can be approximately described by a pure state $|\psi\rangle=C_{g0}|g0\rangle+C_{g1}|g1\rangle+C_{e0}|e0\rangle+C_{e1}|e1\rangle+C_{g2}|g2\rangle$, where $C_{g(e),n}$ are the probability amplitudes for $|g(e),n\rangle$, with $g$ ($e$) the ground (excited) state of the qubit and $n$ the excitation number of the magnon. The steady-state solution of the Schr\"odinger equation $i\partial|\psi\rangle/\partial t=H_{\rm non}|\psi\rangle$ under the condition of $\Delta_+=J$ yields (see the Appendix for details)
\begin{equation}\label{steadysolutions}
\begin{aligned}
C_{g0}&\approx1,\\
C_{e0}&=-\frac{4J(\Omega_m-\Omega_q)+2i\Omega_q\kappa}{\kappa^2+4iJ\kappa},\\
C_{g1}&=-\frac{4J(\Omega_q-\Omega_m)+2i\Omega_m\kappa}{\kappa^2+4iJ\kappa},\\
C_{g2}&=\frac{2\sqrt{2}(A+Bi)}{(\kappa^2-2J^2+4iJ\kappa)(\kappa^2+4iJ\kappa)},\\
C_{e1}&=-\frac{4(C+Di)}{(\kappa^2-2J^2+4iJ\kappa)(\kappa^2+4iJ\kappa)},
\end{aligned}
\end{equation}
where the coefficients $A$, $B$, $C$, and $D$ can be found in Eq.~(\ref{ABCDE}).

\begin{figure}[htbp]
\centering
\includegraphics[width=0.95\linewidth]{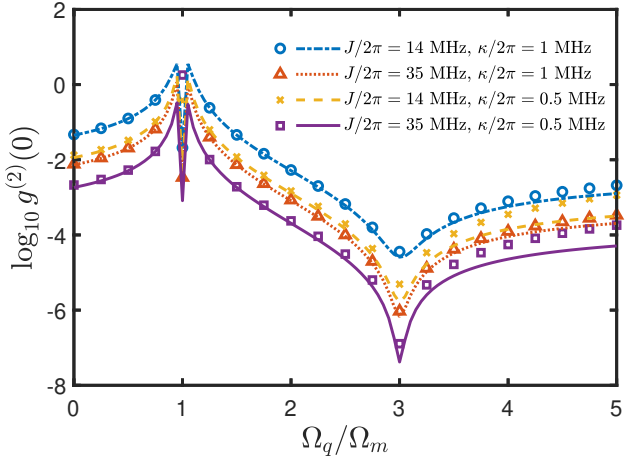}
\caption{Steady-state $\log_{10}g^{(2)}{(0)}$ versus the ratio of the probing field strength to the driving field strength $\Omega_q/\Omega_m$ for $\Delta_+/J=1$. The numerical and the analytical results are indicated by the markers and the lines, respectively. Here, $\Omega_m/2\pi=0.1$ MHz.}\label{comparison}
\end{figure}

With the steady-state amplitudes in Eq.~(\ref{steadysolutions}), the function $g^{(2)}$ can then be expressed by
\begin{equation}\label{g2analytical}
g^{(2)}(0)=\frac{2|C_{g2}|^2}{\left(|C_{g1}|^2+|C_{e1}|^2+2|C_{g2}|^2\right)^2}\approx\frac{2|C_{g2}|^2}{|C_{g1}|^4},
\end{equation}
where the second equivalence is approximately valid for $|C_{g1}|\gg|C_{e1}|, |C_{g2}|$. This result could be verified by the numerical simulation in Fig.~\ref{comparison} for various transversal interaction strengths $J$ and system decay rates $\kappa$. It is demonstrated that for a comparatively strong decay $\kappa/2\pi=1$ MHz, the analytical results in Eq.~(\ref{g2analytical}) for the steady-state correlation function match perfectly with the numerical results by the master equation, while for a comparatively weak decay $\kappa/2\pi=0.5$ MHz, a small deviation appears between them when $\Omega_q/\Omega_m\geq3$. In addition, the numerical results of function $g^{(2)}$ around $\Omega_q=\Omega_m$ are unstable with changing the decay rate. The local minimum point of the function $g^{(2)}$ for $\kappa/2\pi=1$ MHz becomes the local maximum point for $\kappa/2\pi=0.5$ MHz. In contrast, the global minimum point around $\Omega_q=3\Omega_m$ for the strongest magnon antibunching effect is stable. Its presence is independent of individual magnitudes of both $\Omega_q$ and $\Omega_m$. It is consistent with the analysis of the second derivative of the function $g^{(2)}$ in Eq.~(\ref{secondderivative}).

\begin{figure}[htbp]
\centering
\includegraphics[width=0.95\linewidth]{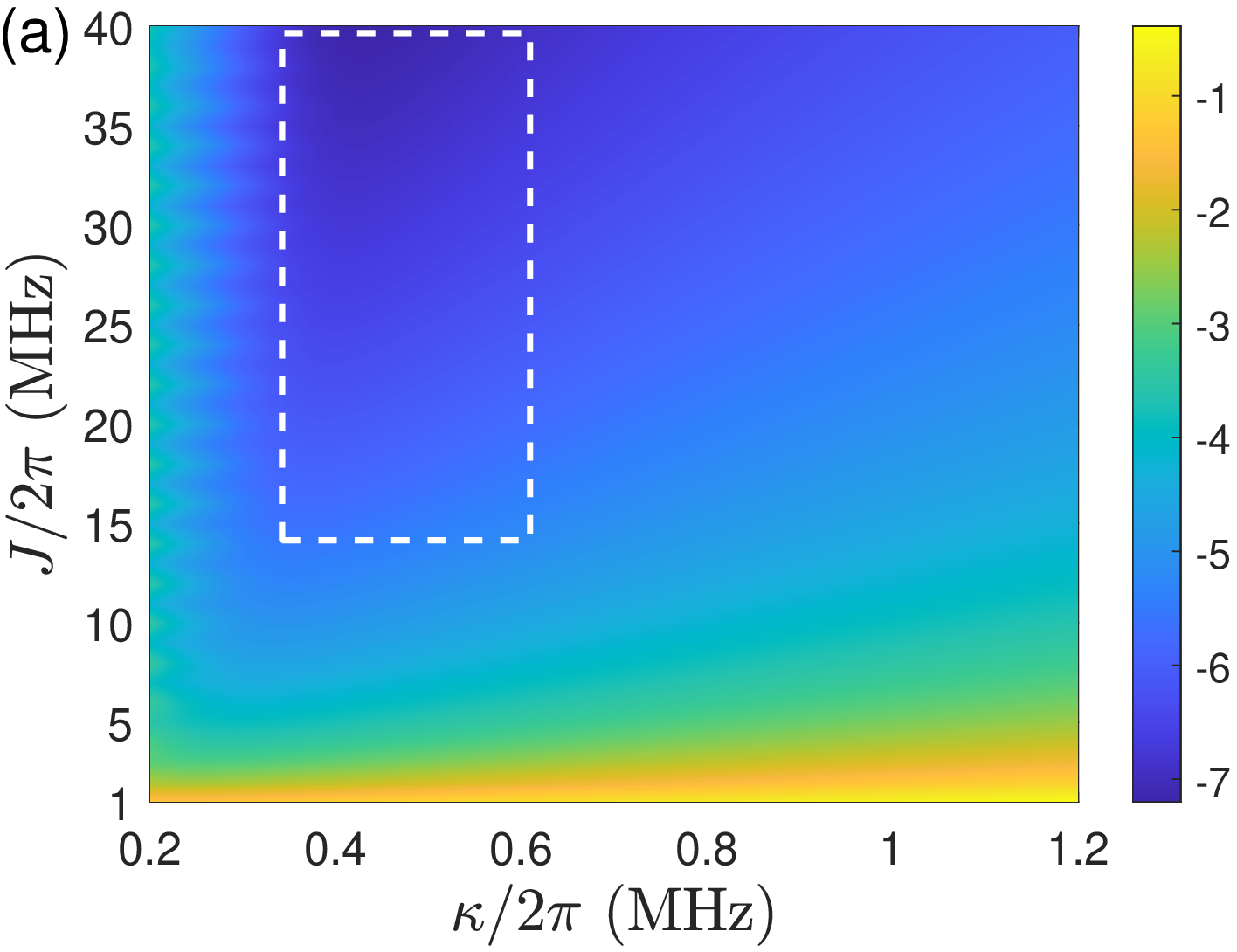}
\includegraphics[width=0.95\linewidth]{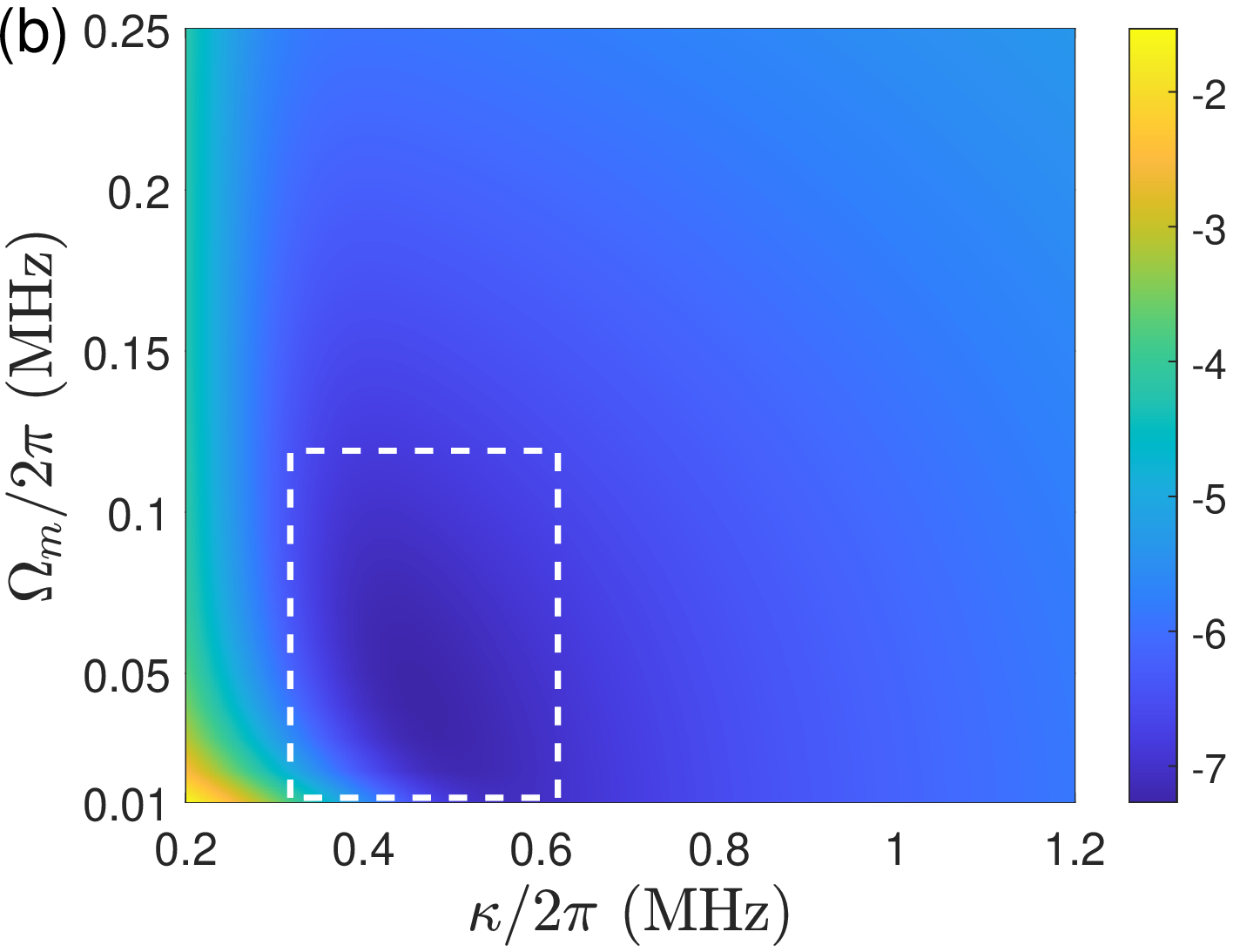}
\caption{Steady-state $\log_{10}g^{(2)}{(0)}$ versus (a) the transversal coupling strength $J$ and (b) the driving field strength $\Omega_m$ and the decay rate $\kappa$ for $\Delta_+=J$ and $\Omega_q=3\Omega_m$. In (a) $\Omega_m/2\pi=0.1$ MHz and in (b) $J/2\pi=35$ MHz.}\label{optdecayOm}
\end{figure}

\begin{figure}[htbp]
\centering
\includegraphics[width=0.95\linewidth]{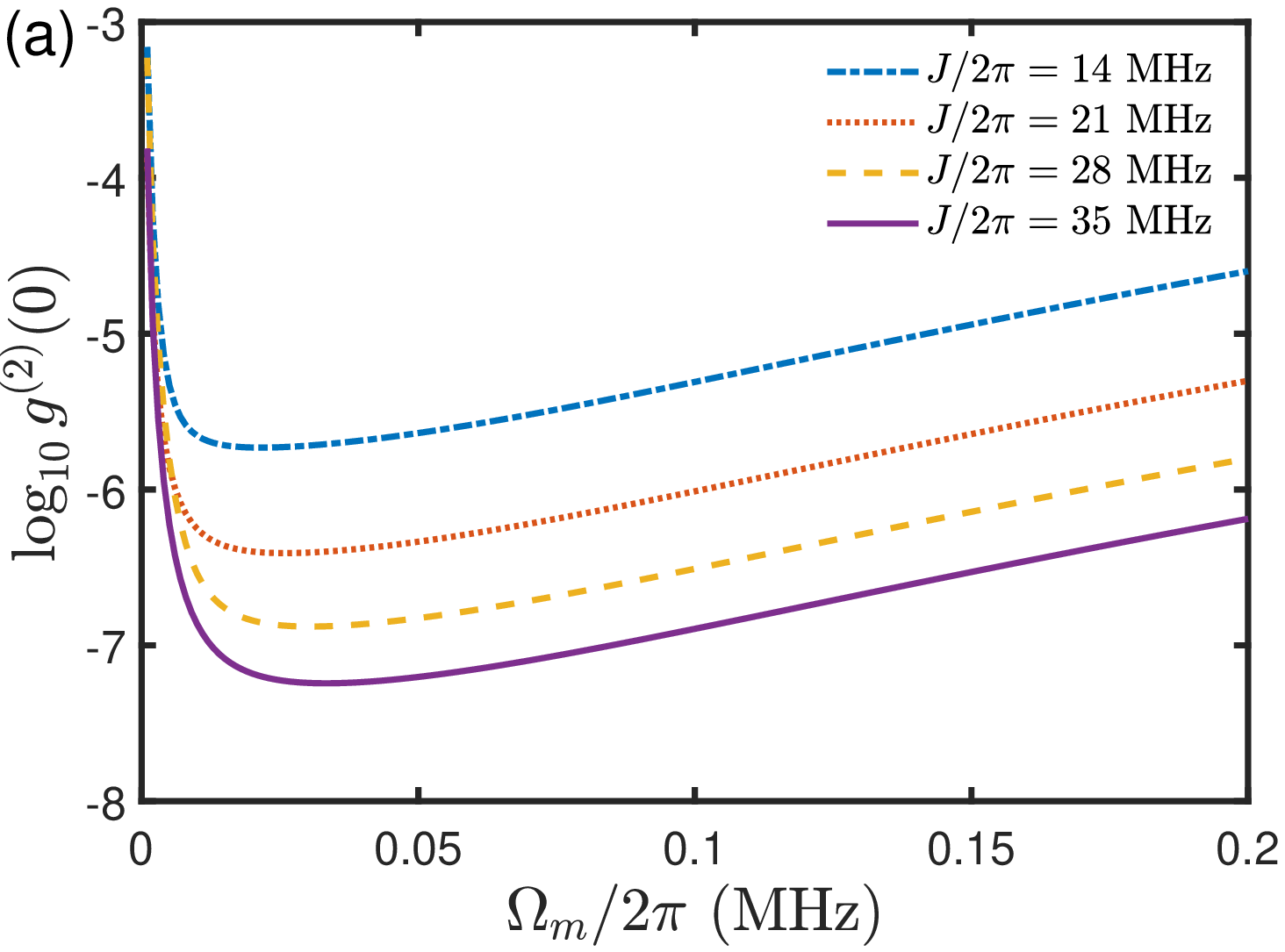}
\includegraphics[width=0.95\linewidth]{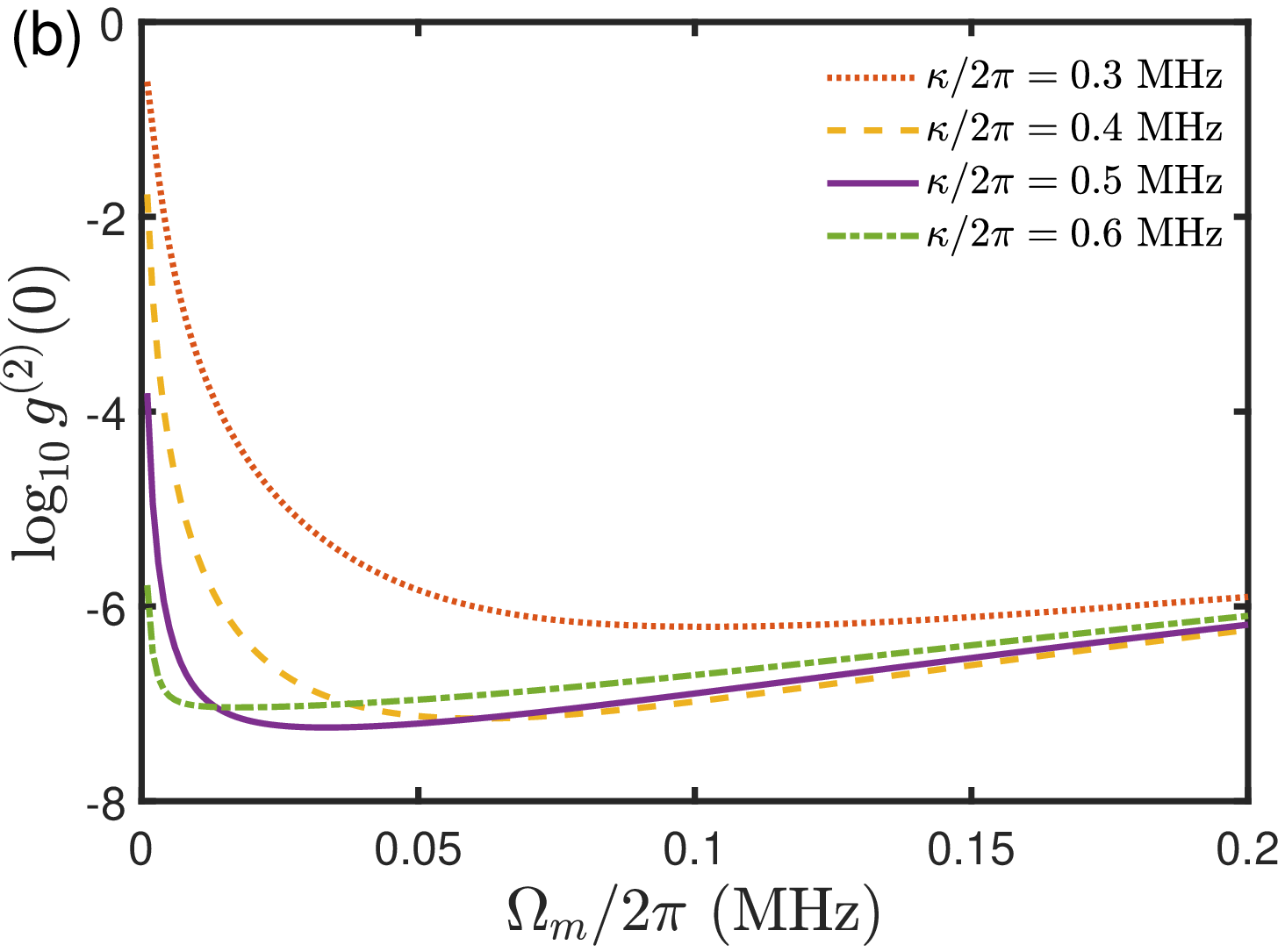}
\caption{Steady-state $\log_{10}g^{(2)}{(0)}$ as a function of (a) the driving field strength $\Omega_m$ and (b) the decay rate $\kappa$, still under the optimized conditions $\Delta_+=J$ and $\Omega_q=3\Omega_m$. In (a) $\kappa/2\pi=0.5$ MHz and in (b) $J/2\pi=35$ MHz.}\label{fixdecay}
\end{figure}

Nevertheless, it seems important to show explicitly that both probing and driving fields are indispensable to enhance magnon blockade. We then explore the optimal connections in Fig.~\ref{optdecayOm} among the transversal coupling strength $J$, optimal driving field strength $\Omega_m$, and decay rate $\kappa$. We are motivated to find the proper intensities of the driving and probing fields for the optimized ratio $\Omega_q=3\Omega_m$. In Fig.~\ref{optdecayOm}(a), with a fixed driving intensity, a region is distinguished by the white dashed lines where the function $g^{(2)}$ exhibits a non-monotonic function of the decay rate and its magnitude is less than $10^{-7}$. We then fix a strong transversal coupling strength $J/2\pi=35$ MHz that is close to the upper bound in present experiments and optimize the function $g^{(2)}$ in the space of $\Omega_m$ and $\kappa$ in Fig.~\ref{optdecayOm}(b). It is interesting to find that there are optimal driving intensity and probing intensity for the best magnon blockade. The region for $g^{(2)}\leq10^{-7}$ is also distinguished by the white dashed lines, where $\Omega_m/2\pi$ ranges from $0.01$ to $0.13$ MHz, $\Omega_q/2\pi$ ranges from $0.03$ to $0.39$ MHz, and $\kappa/2\pi$ ranges from $0.35$ to $0.62$ MHz.

More transparent connections among $\Omega_m$, $J$, and $\kappa$ are demonstrated by Fig.~\ref{fixdecay}. In both Fig.~\ref{fixdecay}(a) under a fixed decay rate and Fig.~\ref{fixdecay}(b) under a fixed transversal coupling strength, the magnon blockade becomes weak and even disappears when the field intensities $\Omega_m$ and $\Omega_q$ approach vanishing. It is found that the function $g^{(2)}$ could be locally minimized by proper $\Omega_m$ and $\Omega_q$. In Fig.~\ref{fixdecay}(a) with $\kappa/2\pi=0.5$ MHz, $\Omega_m$ and $\Omega_q$ for the best blockade become larger for a stronger transversal coupling. In particular, we have $\Omega_m/2\pi=0.021$ MHz and $\Omega_q/2\pi=0.063$ MHz when $J/2\pi=14$ MHz, and $\Omega_m/2\pi=0.033$ MHz and $\Omega_q/2\pi=0.099$ MHz when $J/2\pi=35$ MHz. In Fig.~\ref{fixdecay}(b) with $J/2\pi=35$ MHz, $\Omega_m$ and $\Omega_q$ are optimized with a properly chosen decay rate. When $\kappa/2\pi=0.5$ MHz, a high-degree blockade is achieved with $g^{(2)}(0)=10^{-7.24}$.

In a system consisting of a cavity and a single three-level atom~\cite{Tang2019Strong}, the dressed-state splitting between the higher and lower branches is enhanced by the optical Stark shift. Then the function $g^{(2)}$ can be about three orders lower than that in the JC model [$g^{(2)}(0)\sim10^{-3}$]. In a cavity QED system~\cite{Li2021strong}, both the emitter and cavity are driven by a coherent laser field, and the function $g^{(2)}$ can become as low as $g^{(2)}(0)\sim10^{-4}$. In a double-cavity system~\cite{Wang2020Photon}, the photon blockade with $g^{(2)}(0)\sim10^{-5}$ has been observed. Rather than these systems for photon blockade, the magnon-qubit system is featured with a much stronger transversal coupling and a much faster damping. We can therefore realize a higher degree of blockade [$g^{(2)}(0)\sim10^{-7}$] under the optimized conditions $\Delta_+=J$ and $\Omega_q=3\Omega_m$ with proper driving field strength and system decay rate. Note that our system is established on the direct coupling between the magnon and qubit. In contrast, for the hybrid system on the indirect interaction between the magnon and qubit mediated by virtually eliminating the degrees of freedom of a microwave cavity~\cite{Xie2020Quantum}, the magnon blockade with $g^{(2)}(0)\sim10^{-2}$ can be established through modulating the ratio of the detuning average and the transversal coupling strength.

In experiments, the magnon blockade can be indirectly detected by measuring the magnon number using the atomic Stark shift~\cite{Kounalakis2022Analog}. By virtue of Eq.~(\ref{g2analytical}), the blockade measure $g^{(2)}(0)\equiv\langle m^\dagger m^\dagger mm\rangle/\langle m^\dagger m\rangle^2$ is approximately equivalent to $2P_2/P_1^2$, where $P_i\equiv\langle i|m^\dagger m|i\rangle$ is the population on the number state $|i\rangle$ of the magnon. After the magnon blockade is achieved, one can switch off the driving and probing fields and tune the frequency of either the transom qubit or the magnon mode so that they are far-off-resonance with each other and their effective Hamiltonian becomes $H_{\rm eff}=\omega_m(m^\dagger m+1/2)+\omega_q\sigma_z/2+\chi(m^\dagger m+1/2)\sigma_z$, where $\chi=J^2/|\omega_m-\omega_q|$ is the dispersive coupling strength. In the strong dispersive regime, the transmon qubit serves as a magnon detector, similar to the situation in the cavity-qubit system~\cite{Schuster2007Resolving,Vlastakis2013Deterministically,Vlastakis2015Characterizing,Langford2017Experimentally}. The qubit transition energy can be resolved into a separate spectral line for each magnon number state, and the strength of each line is a measure of the probability of finding the magnon in the relevant number state.

\section{Nonideal situations}\label{discussion}

We have created the magnon blockade in a transversally coupled magnon-qubit system and found its optimal conditions. The preceding calculations of the function $g^{(2)}$ were carried out in a vacuum environment. The external magnetic flux was fixed at a proper value to avoid the longitudinal (radiation-pressure like) interaction between the magnon and qubit. In this section we consider nonideal situations.

\subsection{Thermal noise effects on blockade}\label{thermalnoise}

\begin{figure}[htbp]
\centering
\includegraphics[width=0.95\linewidth]{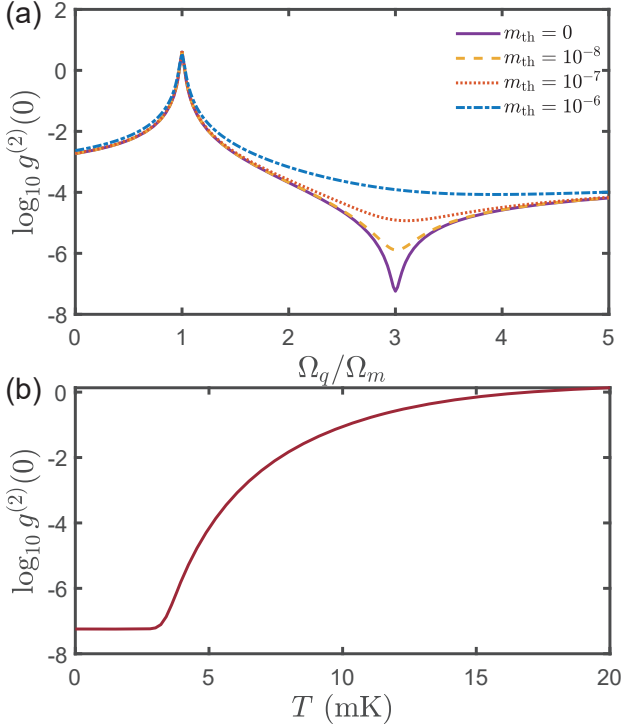}
\caption{(a) Steady-state $\log_{10}g^{(2)}{(0)}$ as a function of the ratio $\Omega_q/\Omega_m$ with various thermal occupation numbers $m_{\rm th}$ of the magnon. (b) Steady-state $\log_{10}g^{(2)}{(0)}$ as a function of the temperature $T$ for $\Omega_q=3\Omega_m$. The other parameters are $\Delta_+=J$, $J/2\pi=35$ MHz, $\Omega_m/2\pi=0.033$ MHz, and $\kappa/2\pi=0.5$ MHz.}\label{Oqnth}
\end{figure}

Taking a nonzero-temperature bosonic environment into consideration, we can discuss the effect of a thermal noise on the magnon blockade. Under the resonant condition $\Delta_-=0$, the master equation~(\ref{masterequation}) becomes
\begin{equation}\label{masterequation_nth}
\begin{aligned}
\frac{\partial}{\partial t}\rho&=-i[H_{\rm eff},\rho]+\frac{\kappa_m}{2}(m_{\rm th}+1)\mathcal{L}_m[\rho]\\
&+\frac{\kappa_m}{2}m_{\rm th}\mathcal{L}_{m^\dagger}[\rho]+\frac{\kappa_q}{2}\mathcal{L}_{\sigma_-}[\rho],\\
\end{aligned}
\end{equation}
where $m_{\rm th}=[\exp(\hbar\omega_m/k_BT)-1]^{-1}$ is the thermal magnon occupation number with the Boltzmann constant $k_B$ and the temperature $T$. The function $g^{(2)}$ is then numerically solved by Eq.~(\ref{masterequation_nth}) with the effective Hamiltonian in Eq.~(\ref{Ham_eff}). In the hybrid magnon-qubit system~\cite{Kounalakis2022Analog}, the direct interaction between the magnon and qubit is established at a cryogenic temperature about $5$ mK~\cite{Kounalakis2022Analog}. With the magnon frequency $1.5$ GHz, the thermal occupation number is approximately $10^{-8}$, $10^{-7}$, and $10^{-6}$ when $T=3.9$,  $T=4.5$, and $T=5.3$ mK, respectively. In Fig.~\ref{Oqnth}(a) with $J/2\pi=35$ MHz and $\Omega_m/2\pi=0.033$ MHz, one can find that the optimal condition around $\Omega_q=3\Omega_m$ can be maintained when $m_{\rm th}<10^{-6}$, although the function $g^{(2)}$ increases with $m_{\rm th}$. In particular, $g^{(2)}(0)=10^{-6.01}$ when $m_{\rm th}=10^{-8}$ and $g^{(2)}(0)=10^{-5.03}$ when $m_{\rm th}=10^{-7}$. With a low decay rate $\kappa/2\pi=0.5$ MHz, a dramatic bunching phenomenon occurs around $\Omega_q=\Omega_m$ for arbitrary $m_{\rm th}$.

In addition, the experimental feasibility of our proposal for magnon blockade is clearly shown in Fig.~\ref{Oqnth}(b) for the dependence of the function $g^{(2)}$ on the environmental temperature. It is interesting to find that the magnon blockade approaches a saturated lowerbound with $g^{(2)}(0)=10^{-7.24}$ when $T<2.6$ mK. In addition, the blockade completely vanishes with $g^{(2)}(0)\ge1$ when $T>17.2$ mK or $m_{\rm th}=0.016$. This result quantifies the restriction from the thermal noise on the magnon blockade.

\subsection{Extra longitudinal interaction}\label{secLongitudinal}

\begin{figure}[htbp]
\centering
\includegraphics[width=0.95\linewidth]{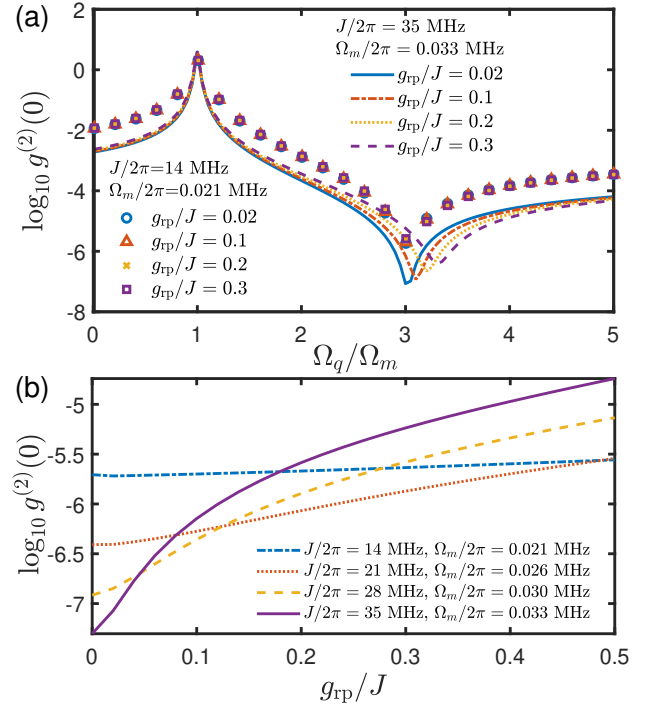}
\caption{(a) Steady-state $\log_{10}g^{(2)}(0)$ as a function of the ratio $\Omega_q/\Omega_m$ with various ratios $g_{\rm rp}/J$. (b) Steady-state $\log_{10}g^{(2)}(0)$ as a function of the ratio $g_{\rm rp}/J$ under the conditions of $\Omega_q=3\Omega_m$. The other parameters are $\Delta_+=J$, $\omega/2\pi=1.5$ GHz, and $\kappa/2\pi=0.5$ MHz.}\label{longitudinalfig}
\end{figure}

In this section we discuss the effect of the extra magnon-qubit longitudinal interaction $g_{\rm rp}\sigma_+\sigma_-(m+m^\dagger)$ on the magnon blockade. The coupling strength $g_{\rm rp}$ can be modulated by the external flux $\Phi_b$ on qubit~\cite{Kounalakis2022Analog}. In the rotating frame with respect to $H'=\omega(m^\dagger m+\sigma_+\sigma_-)$, the effective time-dependent Hamiltonian becomes
\begin{equation}\label{Ham_longitudinal}
\begin{aligned}
H_{\rm eff}&=\Delta_+\sigma_+\sigma_-+\Delta_+m^\dagger m+J(m\sigma_++m^\dagger\sigma_-)\\
&+g_{\rm rp}\sigma_+\sigma_-(me^{-i\omega t}+m^\dagger e^{i\omega t})\\
&+\Omega_m(m^\dagger+m)+\Omega_q(\sigma_++\sigma_-).
\end{aligned}
\end{equation}
Formally, this model is indeed a hybrid of cavity QED and cavity optomechanics for blockade.

Using the master equation~(\ref{masterequation}) with the effective Hamiltonian $H_{\rm eff}$ in Eq.~(\ref{Ham_longitudinal}), we plot the steady-state function $g^{(2)}$ in Fig.~\ref{longitudinalfig}(a) versus the ratio $\Omega_q/\Omega_m$ for various ratios $g_{\rm rp}/J$. We use the optimal driving intensity $\Omega_m$ presented in Fig.~\ref{fixdecay}(a) for different $J$. It is interesting to find that under a sufficiently large $J$, the optimal ratio $\Omega_q/\Omega_m$ deviates gradually from $\Omega_q/\Omega_m=3$ and the expectation value of the function $g^{(2)}$ becomes even larger by increasing the extra longitudinal interaction. In particular, when the transversal coupling is as strong as $J/2\pi=35$ MHz, the optimized ratio of driving and probing fields becomes about $\Omega_q/\Omega_m\approx3.1$ with $g^{(2)}(0)=10^{-6.93}$ for $g_{\rm rp}/J=0.1$, $\Omega_q/\Omega_m\approx3.2$ with $g^{(2)}(0)=10^{-6.65}$ for $g_{\rm rp}/J=0.2$, and $\Omega_q/\Omega_m\approx3.3$ with $g^{(2)}(0)=10^{-6.40}$ for $g_{\rm rp}/J=0.3$. In the weak coupling region for $J$, a nonvanishing $g_{\rm rp}$ has no significant impact on the function $g^{(2)}$ for arbitrary $\Omega_q/\Omega_m$. In particular, when $J/2\pi=14$ MHz, the optimal condition $\Omega_q/\Omega_m=3$ remains unchanged even if the extra longitudinal interaction is about $g_{\rm rp}/J=0.3$.

Moreover, we plot the steady-state function $g^{(2)}$ in Fig.~\ref{longitudinalfig}(b) under the optimized conditions of $\Delta_+=J$ and $\Omega_q=3\Omega_m$ with proper transversal coupling strengths and driving intensities that minimize the function $g^{(2)}$ in Fig.~\ref{fixdecay}(a). It is confirmed that the sensitivity of the equal-time correlation function to the ratio $g_{\rm rp}/J$ is enhanced by increasing the transversal coupling strength $J$. In Fig.~\ref{longitudinalfig}(b), for a weak transversal interaction with $J/2\pi=14$ MHz, the function $g^{(2)}$ is nearly invariant in the presence of the extra longitudinal interaction. When $g_{\rm rp}/J<0.04$, it is found that the degree of magnon blockade can be enhanced by increasing the transversal coupling strength $J$. However, the function $g^{(2)}$ for $J/2\pi=35$ MHz becomes larger than those for $J/2\pi=28$, $21$, and $14$ MHz, when $g_{\rm rp}/J$ approaches $0.04$, $0.08$, and $0.18$, respectively. The expectation value of the function $g^{(2)}$ is enhanced by almost three orders of magnitude when $g_{\rm rp}/J$ increases from zero to $0.5$. This indicates that the magnon blockade can no longer be promoted by increasing the transversal interaction in the presence of a significant longitudinal interaction. In contrast, with a moderate $J$, e.g., $J/2\pi=21$ MHz, we can achieve $g^{(2)}(0)\sim 10^{-6}$ even when $g_{\rm rp}/J=0.3$. Our magnon-blockade proposal, as well as the optimal condition for $\Omega_q/\Omega_m$, is therefore robust in the presence of the extra longitudinal interaction. Nevertheless, the combination of cavity QED and cavity optomechanics is not favorable to blockade.

\section{Conclusion}\label{Conclusion}

In summary, we proposed to generate and optimize the magnon blockade in a system established by the direct interaction between a YIG sphere and a superconducting transmon qubit. With a weak driving field on the magnon, a weak probing field on the qubit, a strong transversal magnon-qubit interaction, and an extremely low damping rate of the magnon, a high-degree magnon blockade could be achieved by minimizing the steady-state second-order correlation function. It was found by both analytical derivation and numerical simulation that $g^{(2)}(0)\sim10^{-7}$ is available with the parameters in current experiments, when the transversal coupling strength is equivalent to the average detuning of the magnon and qubit and the driving-field intensity is about three times the probing-field intensity. The function $g^{(2)}$ for the best magnon blockade is about two orders of magnitude lower than that for photon blockade due to the unaccessible regimes in cavity QED and cavity optomechanics about the strong interaction and the low decay rate. We also discussed the influence of the environmental temperature and the extra longitudinal interaction between the magnon and qubit on the magnon blockade. Our proposal illustrates a systematic optimization process for magnon blockade. It paves a way towards the on-chip magnonics in the nonclassical limit and will facilitate the preparation of a single magnon state and even an arbitrary superposition of magnon Fock states.

\section*{Acknowledgments}

We acknowledge financial support from the National Natural Science Foundation of China (Grant No. 11974311).

\appendix

\section{Analytical model of optimized conditions for blockade}\label{Analytic model}

This appendix contributes to confirming the analytical expression for the steady-state equal-time second-order correlation function in Eq.~(\ref{g2analytical}), identifying the optimized point for blockade in the ratio of $\Omega_q/\Omega_m$, and providing the population dynamics of the magnon mode by numerical simulation to strengthen the evidence of magnon blockade.

According to the analytical model in Ref.~\cite{Kar2013Single}, the non-Hermitian Schr\"odinger equation $i\partial|\psi\rangle/\partial t=H_{\rm non}|\psi\rangle$ is consistent with the master equation~(\ref{masterequation}) under the no-quantum-jump assumption, where $|\psi\rangle=C_{g0}|g0\rangle+C_{g1}|g1\rangle+C_{e0}|e0\rangle+C_{e1}|e1\rangle+C_{g2}|g2\rangle$ and $H_{\rm non}=H_{\rm eff}-i\kappa(\sigma_+\sigma_-+m^\dagger m)/2$. Consequently, we have
\begin{equation}\label{differential}
\begin{aligned}
i\dot{C}_{g0}&=0,\\
i\dot{C}_{e0}&=JC_{g1}+\Omega_qC_{g0}+(\Delta_+-i\kappa/2)C_{e0},\\
i\dot{C}_{g1}&=JC_{e0}+\Omega_mC_{g0}+(\Delta_+-i\kappa/2)C_{g1},\\
i\dot{C}_{e1}&=\sqrt{2}JC_{g2}+\Omega_qC_{g1}+\Omega_mC_{e0}+2(\Delta_+-i\kappa/2)C_{e1},\\
i\dot{C}_{g2}&=\sqrt{2}JC_{e1}+\sqrt{2}\Omega_mC_{g1}+2(\Delta_+-i\kappa/2)C_{g2}.
\end{aligned}
\end{equation}
Then its steady-state solution can be obtained from
\begin{equation}\label{differentialsteady}
\begin{aligned}
0&=JC_{g1}+\Omega_qC_{g0}+(\Delta_+-i\kappa/2)C_{e0},\\
0&=JC_{e0}+\Omega_mC_{g0}+(\Delta_+-i\kappa/2)C_{g1},\\
0&=\sqrt{2}JC_{g2}+\Omega_qC_{g1}+\Omega_mC_{e0}+2(\Delta_+-i\kappa/2)C_{e1},\\
0&=\sqrt{2}JC_{e1}+\sqrt{2}\Omega_mC_{g1}+2(\Delta_+-i\kappa/2)C_{g2}.
\end{aligned}
\end{equation}
Under $\Delta_+=J$, which is the optimal condition confirmed in Fig.~\ref{conditionDelta}, we have
\begin{equation}\label{steadysolutions_app}
\begin{aligned}
C_{g0}&\approx1\\
C_{e0}&=-\frac{4J(\Omega_m-\Omega_q)+2i\Omega_q\kappa}{\kappa^2+4iJ\kappa},\\
C_{g1}&=-\frac{4J(\Omega_q-\Omega_m)+2i\Omega_m\kappa}{\kappa^2+4iJ\kappa},\\
C_{g2}&=\frac{2\sqrt{2}(A+Bi)}{(\kappa^2-2J^2+4iJ\kappa)(\kappa^2+4iJ\kappa)},\\
C_{e1}&=-\frac{4(C+Di)}{(\kappa^2-2J^2+4iJ\kappa)(\kappa^2+4iJ\kappa)},
\end{aligned}
\end{equation}
where
\begin{equation}\label{ABCDE}
\begin{aligned}
A&=2J^2(3\Omega_m^2+\Omega_q^2-4\Omega_m\Omega_q)-\Omega_m^2\kappa^2,\\
B&=4J\kappa(\Omega_m\Omega_q-\Omega_m^2),\\
C&=2J^2(2\Omega_m^2+\Omega_q^2-3\Omega_m\Omega_q)+\Omega_m\Omega_q\kappa^2,\\
D&=J\kappa(4\Omega_m\Omega_q-2\Omega_m^2-\Omega_q^2).
\end{aligned}
\end{equation}
With the steady amplitudes in Eq.~(\ref{steadysolutions_app}), the function $g^{(2)}$ defined in Eq.~(\ref{g2define}) can be expressed by
\begin{equation}\label{g2analyticalDetailed}
\begin{aligned}
&g^{(2)}(0)=\frac{2|C_{g2}|^2}{(|C_{g1}|^2+|C_{e1}|^2+2|C_{g2}|^2)^2}\approx\frac{2|C_{g2}|^2}{|C_{g1}|^4}\\
&=\frac{(A^2+B^2)(\kappa^4+16J^2\kappa^2)}{[(2J^2-\kappa^2)^2+16J^2\kappa^2][4J^2(\Omega_q-\Omega_m)^2+\Omega_m^2\kappa^2]^2},
\end{aligned}
\end{equation}
which is valid for $|C_{g1}|\gg|C_{e1}|, |C_{g2}|$. Using $\Omega_q=(l+1)\Omega_m$ and $r=\kappa/J$, we can rewrite the function $g^{(2)}$ in Eq.~(\ref{g2analyticalDetailed}) as a function of the dimensionless parameters $l$ and $r$:
\begin{equation}\label{g2analyticalDetailed_n}
g^{(2)}(0)=\frac{4(l-2)^2l^2+4r^2(3l+2)l+r^4}{\left[1+\frac{4(1-r^2)}{r^4+16r^2}\right](4l^2+r^2)^2}.
\end{equation}
The first derivative of $g^{(2)}(0)$ with respect to $l$ is
\begin{equation}\label{g2analyticalderive}
\begin{aligned}
\frac{dg^{(2)}(0)}{dl}&=\frac{l^4+bl^3+cl^2+dl+f}{\left[1+\frac{4(1-r^2)}{r^4+16r^2}\right](l^2+\frac{r^2}{4})^3},
\end{aligned}
\end{equation}
where $b=-5r^2/4-2$, $c=-9r^2/4$, $d=r^4/8+r^2/2$, and $f=r^4/8$. With the Galois theory~\cite{Harold1984Galois}, the numerator of Eq.~(\ref{g2analyticalderive}) can be reorganized into
\begin{equation}\label{reducedsteptwo}
\left(l^2+\frac{1}{2}bl+\frac{1}{2}y\right)^2-(l-l')^2,
\end{equation}
where
\begin{equation}
\begin{aligned}
l'&=\frac{2d-by}{b^2-4c+4y},\\
y&=\frac{c}{3}+\frac{-1-\sqrt{3}i}{12}\left(\sqrt{h_1}+\sqrt{h_1+h_2}\right)^{1/3}\\
&+\frac{-1+\sqrt{3}i}{12}\left(\sqrt{h_1}-\sqrt{h_1+h_2}\right)^{1/3}
\end{aligned}
\end{equation}
with $h_1=(-36cc_1+8c^3-108d_1)^2$, $h_2=(12c_1-4c^2)^3$, $c_1=-(4f-bd)$, and $d_1=-(b^2-4c)f-d^2$. Then we can find two real roots of $l=\Omega_q/\Omega_m-1$ for $dg^{(2)}(0)/dl=0$,
\begin{equation}\label{roots}
\begin{aligned}
l_1&=\frac{-\frac{1}{2}b+1+\sqrt{\frac{b^2}{4}-b-2y-4l'+1}}{2},\\
l_2&=\frac{-\frac{1}{2}b+1-\sqrt{\frac{b^2}{4}-b-2y-4l'+1}}{2},
\end{aligned}
\end{equation}
which are larger than $-1$ in the positive region of $r$.

\begin{figure}[htbp]
\centering
\includegraphics[width=0.95\linewidth]{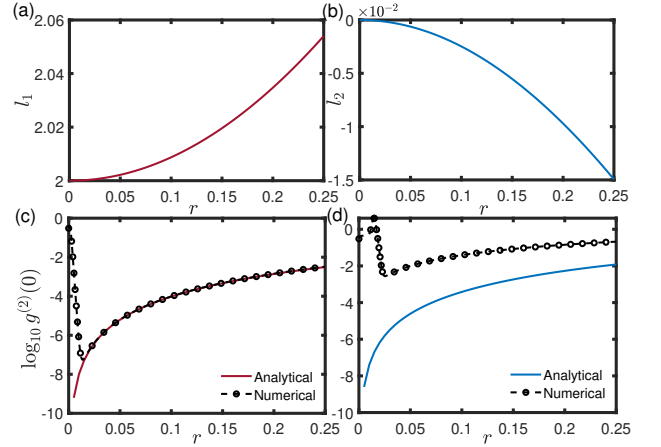}
\caption{Plots of (a) $l_1$ and (b) $l_2$ versus the ratio $r$ of the system decay rate and the transversal coupling strength. (c) and (d): Analytical [Eq.~(\ref{g2analyticalDetailed_n})] and numerical results [Eq.~(\ref{masterequation})] for the steady-state $\log_{10}g^{(2)}{(0)}$ versus the ratio $r$ with the roots $l_1$ and $l_2$, respectively. The other parameters are $\Delta_+=J$, $J/2\pi=35$ MHz, and $\Omega_m/2\pi=0.033$ MHz.}\label{t1t2}
\end{figure}

\begin{figure}[htbp]
\centering
\includegraphics[width=0.95\linewidth]{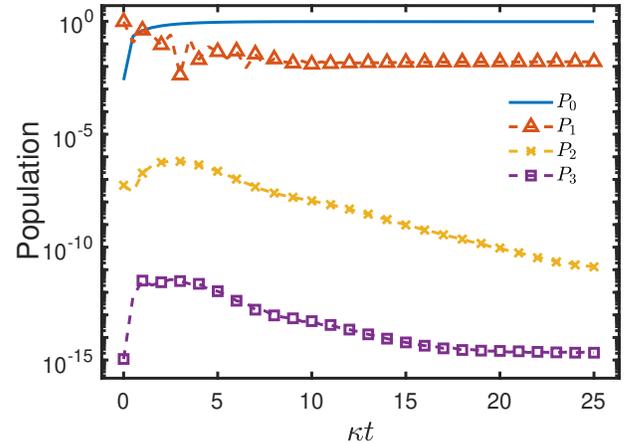}
\caption{Population dynamics of a few lowest Fock states of the magnon mode under the optimal conditions of $\Delta_+=J$ and $\Omega_q=3\Omega_m$. The other parameters are $J/2\pi=35$ MHz, $\Omega_m/2\pi=0.033$ MHz, and $\kappa/2\pi=0.5$ MHz. The initial state is set as $\rho(0)=|g1\rangle\langle g1|$.}\label{density}
\end{figure}

Here $l$ is a function uniquely determined by $r$, i.e., the ratio of the system decay rate and the transversal coupling strength $\kappa/J$. In Figs.~\ref{t1t2}(a) and \ref{t1t2}(b) we can find the two extremum points $l_1\approx2$ and $l_2\approx0$ within a range of experimentally relevant parameter $r\in(0, 0.25)$. We plot both the analytical and numerical results for the function $g^{(2)}$ in Figs.~\ref{t1t2}(c) and \ref{t1t2}(d), respectively. It is found in Fig.~\ref{t1t2}(c) that when $r\ge0.014$, the numerical simulation could be perfectly interpreted by the analytical evaluation around the root $l_1\approx2$. In particular, the turning point of $r\approx0.014$ corresponds to the purple line in Fig.~\ref{fixdecay}(b) for $\kappa/2\pi=0.5$ MHz. The analytical model for blockade becomes invalid in the presence of a sufficiently small decay rate. It is thus reasonable to understand the derivation between the numerical simulation and analytical evaluation when $r$ becomes smaller than the turning point. In contrast, it is found in Fig.~\ref{t1t2}(d) that the analytical model around the root $l_2\approx0$ is unable to interpret the numerical results in quantity. The turning point in Fig.~\ref{t1t2}(d), i.e., $r\approx0.02$, corresponds to $\kappa/2\pi\approx0.7$ MHz, predicting the disappearance of the local minimum point around $\Omega_q=\Omega_m$ in Figs.~\ref{comparison}, \ref{Oqnth}(a), and \ref{longitudinalfig}(a). Also, the magnitude of the function $g^{(2)}$ around $l_2$ is much higher than that around $l_1$ for the same $r$.

The optimized condition $l_1\approx2$ or $\Omega_q/\Omega_m\approx3$ can be further confirmed by the second derivative of the function $g^{(2)}$ with respect to $l$,
\begin{equation}\label{secondderivative}
\frac{d^2[g^{(2)}(0)]}{dl^2}=\frac{-2l^5+b'l^4+c'l^3+d'l^2+f'l+g'}
{\left[1+\frac{4(1-r^2)}{r^4+16r^2}\right](l^2+\frac{r^2}{4})^4},
\end{equation}
where $b'=-3b$, $c'=-4c+r^2$, $d'=-5d+9r^2$, $f'=cr^2/2-6f$, and $g'=dr^2/4$. We have $d^2[g^{(2)}(0)]/dl^2>0$ for $l_1\approx2$ and $r\in(0, 0.25)$. It is therefore found that the function $g^{(2)}$ takes a stable and global minimum value around $\Omega_q=3\Omega_m$.

Our protocol for magnon blockade is to explore a theoretical method to push the magnon system to a state with a significant occupation on the first excited level and a nearly vanishing occupation on the other excited levels, which yields a minimum value of the function $g^{(2)}$. In fact, we have $g^{(2)}(0)\equiv\langle m^\dagger m^\dagger mm\rangle/\langle m^\dagger m\rangle^2\approx2P_2/P_1^2$, which is consistent with Eq.~(\ref{g2analytical}) or ~(\ref{g2analyticalDetailed}). In Fig.~\ref{density} we show the population dynamics of a few of the lowest Fock states of the magnon mode under the optimal conditions. It is found that the population on the single-excitation level $P_1$ is about $10^{-2}$ and it is about nine orders larger than that on the double-excitation level $P_2$. In Ref.~\cite{Wang2020Photon} for almost the best photon blockade, where $g^{(2)}(0)\approx10^{-5}$, it turns out that $P_1\leq10^{-2}$ and $P_2\approx10^{-9}$, which is then a strong evidence of the magnon blockade and determines further control in the quantum regime.

\bibliographystyle{apsrevlong}
\bibliography{ref}

\begin{thebibliography}{75}%
\makeatletter
\providecommand \@ifxundefined [1]{%
 \@ifx{#1\undefined}
}%
\providecommand \@ifnum [1]{%
 \ifnum #1\expandafter \@firstoftwo
 \else \expandafter \@secondoftwo
 \fi
}%
\providecommand \@ifx [1]{%
 \ifx #1\expandafter \@firstoftwo
 \else \expandafter \@secondoftwo
 \fi
}%
\providecommand \natexlab [1]{#1}%
\providecommand \enquote  [1]{``#1''}%
\providecommand \bibnamefont  [1]{#1}%
\providecommand \bibfnamefont [1]{#1}%
\providecommand \citenamefont [1]{#1}%
\providecommand \href@noop [0]{\@secondoftwo}%
\providecommand \href [0]{\begingroup \@sanitize@url \@href}%
\providecommand \@href[1]{\@@startlink{#1}\@@href}%
\providecommand \@@href[1]{\endgroup#1\@@endlink}%
\providecommand \@sanitize@url [0]{\catcode `\\12\catcode `\$12\catcode
  `\&12\catcode `\#12\catcode `\^12\catcode `\_12\catcode `\%12\relax}%
\providecommand \@@startlink[1]{}%
\providecommand \@@endlink[0]{}%
\providecommand \url  [0]{\begingroup\@sanitize@url \@url }%
\providecommand \@url [1]{\endgroup\@href {#1}{\urlprefix }}%
\providecommand \urlprefix  [0]{URL }%
\providecommand \Eprint [0]{\href }%
\providecommand \doibase [0]{http://dx.doi.org/}%
\providecommand \selectlanguage [0]{\@gobble}%
\providecommand \bibinfo  [0]{\@secondoftwo}%
\providecommand \bibfield  [0]{\@secondoftwo}%
\providecommand \translation [1]{[#1]}%
\providecommand \BibitemOpen [0]{}%
\providecommand \bibitemStop [0]{}%
\providecommand \bibitemNoStop [0]{.\EOS\space}%
\providecommand \EOS [0]{\spacefactor3000\relax}%
\providecommand \BibitemShut  [1]{\csname bibitem#1\endcsname}%
\let\auto@bib@innerbib\@empty
%</preamble>
\bibitem [{\citenamefont {Wallquist}\ \emph {et~al.}(2009)\citenamefont
  {Wallquist}, \citenamefont {Hammerer}, \citenamefont {Rabl}, \citenamefont
  {Lukin},\ and\ \citenamefont {Zoller}}]{Wallquist2009Hybrid}%
  \BibitemOpen
  \bibfield  {author} {\bibinfo {author} {\bibfnamefont {M.}~\bibnamefont
  {Wallquist}}, \bibinfo {author} {\bibfnamefont {K.}~\bibnamefont {Hammerer}},
  \bibinfo {author} {\bibfnamefont {P.}~\bibnamefont {Rabl}}, \bibinfo {author}
  {\bibfnamefont {M.}~\bibnamefont {Lukin}}, \ and\ \bibinfo {author}
  {\bibfnamefont {P.}~\bibnamefont {Zoller}},\ }\bibfield  {title} {\emph
  {\bibinfo {title} {Hybrid quantum devices and quantum engineering},\ }}\href
  {\doibase 10.1088/0031-8949/2009/T137/014001} {\bibfield  {journal} {\bibinfo
   {journal} {Phys. Scr.}\ }\textbf {\bibinfo {volume} {T137}},\ \bibinfo
  {pages} {014001} (\bibinfo {year} {2009})}\BibitemShut {NoStop}%
\bibitem [{\citenamefont {Kimble}(2008)}]{Kimble2008internet}%
  \BibitemOpen
  \bibfield  {author} {\bibinfo {author} {\bibfnamefont {H.~J.}\ \bibnamefont
  {Kimble}},\ }\bibfield  {title} {\emph {\bibinfo {title} {The quantum
  internet},\ }}\href {\doibase 10.1038/nature07127} {\bibfield  {journal}
  {\bibinfo  {journal} {Nature (London)}\ }\textbf {\bibinfo {volume} {453}},\
  \bibinfo {pages} {1023} (\bibinfo {year} {2008})}\BibitemShut {NoStop}%
\bibitem [{\citenamefont {Imamo\ifmmode~\breve{g}\else
  \u{g}\fi{}lu}(2009)}]{Imamo2009Cavity}%
  \BibitemOpen
  \bibfield  {author} {\bibinfo {author} {\bibfnamefont {A.}~\bibnamefont
  {Imamo\ifmmode~\breve{g}\else \u{g}\fi{}lu}},\ }\bibfield  {title} {\emph
  {\bibinfo {title} {Cavity {QED} based on collective magnetic dipole coupling:
  Spin ensembles as hybrid two-level systems},\ }}\href {\doibase
  10.1103/PhysRevLett.102.083602} {\bibfield  {journal} {\bibinfo  {journal}
  {Phys. Rev. Lett.}\ }\textbf {\bibinfo {volume} {102}},\ \bibinfo {pages}
  {083602} (\bibinfo {year} {2009})}\BibitemShut {NoStop}%
\bibitem [{\citenamefont {Verd\'u}\ \emph {et~al.}(2009)\citenamefont
  {Verd\'u}, \citenamefont {Zoubi}, \citenamefont {Koller}, \citenamefont
  {Majer}, \citenamefont {Ritsch},\ and\ \citenamefont
  {Schmiedmayer}}]{Verd2009Strong}%
  \BibitemOpen
  \bibfield  {author} {\bibinfo {author} {\bibfnamefont {J.}~\bibnamefont
  {Verd\'u}}, \bibinfo {author} {\bibfnamefont {H.}~\bibnamefont {Zoubi}},
  \bibinfo {author} {\bibfnamefont {C.}~\bibnamefont {Koller}}, \bibinfo
  {author} {\bibfnamefont {J.}~\bibnamefont {Majer}}, \bibinfo {author}
  {\bibfnamefont {H.}~\bibnamefont {Ritsch}}, \ and\ \bibinfo {author}
  {\bibfnamefont {J.}~\bibnamefont {Schmiedmayer}},\ }\bibfield  {title} {\emph
  {\bibinfo {title} {Strong magnetic coupling of an ultracold gas to a
  superconducting waveguide cavity},\ }}\href {\doibase
  10.1103/PhysRevLett.103.043603} {\bibfield  {journal} {\bibinfo  {journal}
  {Phys. Rev. Lett.}\ }\textbf {\bibinfo {volume} {103}},\ \bibinfo {pages}
  {043603} (\bibinfo {year} {2009})}\BibitemShut {NoStop}%
\bibitem [{\citenamefont {Eddins}\ \emph {et~al.}(2014)\citenamefont {Eddins},
  \citenamefont {Beedle}, \citenamefont {Hendrickson},\ and\ \citenamefont
  {Friedman}}]{Eddins2014collective}%
  \BibitemOpen
  \bibfield  {author} {\bibinfo {author} {\bibfnamefont {A.~W.}\ \bibnamefont
  {Eddins}}, \bibinfo {author} {\bibfnamefont {C.~C.}\ \bibnamefont {Beedle}},
  \bibinfo {author} {\bibfnamefont {D.~N.}\ \bibnamefont {Hendrickson}}, \ and\
  \bibinfo {author} {\bibfnamefont {J.~R.}\ \bibnamefont {Friedman}},\
  }\bibfield  {title} {\emph {\bibinfo {title} {Collective coupling of a
  macroscopic number of single-molecule magnets with a microwave cavity mode},\
  }}\href {\doibase 10.1103/PhysRevLett.112.120501} {\bibfield  {journal}
  {\bibinfo  {journal} {Phys. Rev. Lett.}\ }\textbf {\bibinfo {volume} {112}},\
  \bibinfo {pages} {120501} (\bibinfo {year} {2014})}\BibitemShut {NoStop}%
\bibitem [{\citenamefont {Kubo}\ \emph {et~al.}(2010)\citenamefont {Kubo},
  \citenamefont {Ong}, \citenamefont {Bertet}, \citenamefont {Vion},
  \citenamefont {Jacques}, \citenamefont {Zheng}, \citenamefont {Dr\'eau},
  \citenamefont {Roch}, \citenamefont {Auffeves}, \citenamefont {Jelezko},
  \citenamefont {Wrachtrup}, \citenamefont {Barthe}, \citenamefont {Bergonzo},\
  and\ \citenamefont {Esteve}}]{Kubo2010Strong}%
  \BibitemOpen
  \bibfield  {author} {\bibinfo {author} {\bibfnamefont {Y.}~\bibnamefont
  {Kubo}}, \bibinfo {author} {\bibfnamefont {F.~R.}\ \bibnamefont {Ong}},
  \bibinfo {author} {\bibfnamefont {P.}~\bibnamefont {Bertet}}, \bibinfo
  {author} {\bibfnamefont {D.}~\bibnamefont {Vion}}, \bibinfo {author}
  {\bibfnamefont {V.}~\bibnamefont {Jacques}}, \bibinfo {author} {\bibfnamefont
  {D.}~\bibnamefont {Zheng}}, \bibinfo {author} {\bibfnamefont
  {A.}~\bibnamefont {Dr\'eau}}, \bibinfo {author} {\bibfnamefont {J.-F.}\
  \bibnamefont {Roch}}, \bibinfo {author} {\bibfnamefont {A.}~\bibnamefont
  {Auffeves}}, \bibinfo {author} {\bibfnamefont {F.}~\bibnamefont {Jelezko}},
  \bibinfo {author} {\bibfnamefont {J.}~\bibnamefont {Wrachtrup}}, \bibinfo
  {author} {\bibfnamefont {M.~F.}\ \bibnamefont {Barthe}}, \bibinfo {author}
  {\bibfnamefont {P.}~\bibnamefont {Bergonzo}}, \ and\ \bibinfo {author}
  {\bibfnamefont {D.}~\bibnamefont {Esteve}},\ }\bibfield  {title} {\emph
  {\bibinfo {title} {Strong coupling of a spin ensemble to a superconducting
  resonator},\ }}\href {\doibase 10.1103/PhysRevLett.105.140502} {\bibfield
  {journal} {\bibinfo  {journal} {Phys. Rev. Lett.}\ }\textbf {\bibinfo
  {volume} {105}},\ \bibinfo {pages} {140502} (\bibinfo {year}
  {2010})}\BibitemShut {NoStop}%
\bibitem [{\citenamefont {Ams\"uss}\ \emph {et~al.}(2011)\citenamefont
  {Ams\"uss}, \citenamefont {Koller}, \citenamefont {N\"obauer}, \citenamefont
  {Putz}, \citenamefont {Rotter}, \citenamefont {Sandner}, \citenamefont
  {Schneider}, \citenamefont {Schramb\"ock}, \citenamefont {Steinhauser},
  \citenamefont {Ritsch}, \citenamefont {Schmiedmayer},\ and\ \citenamefont
  {Majer}}]{Ams2011Cavity}%
  \BibitemOpen
  \bibfield  {author} {\bibinfo {author} {\bibfnamefont {R.}~\bibnamefont
  {Ams\"uss}}, \bibinfo {author} {\bibfnamefont {C.}~\bibnamefont {Koller}},
  \bibinfo {author} {\bibfnamefont {T.}~\bibnamefont {N\"obauer}}, \bibinfo
  {author} {\bibfnamefont {S.}~\bibnamefont {Putz}}, \bibinfo {author}
  {\bibfnamefont {S.}~\bibnamefont {Rotter}}, \bibinfo {author} {\bibfnamefont
  {K.}~\bibnamefont {Sandner}}, \bibinfo {author} {\bibfnamefont
  {S.}~\bibnamefont {Schneider}}, \bibinfo {author} {\bibfnamefont
  {M.}~\bibnamefont {Schramb\"ock}}, \bibinfo {author} {\bibfnamefont
  {G.}~\bibnamefont {Steinhauser}}, \bibinfo {author} {\bibfnamefont
  {H.}~\bibnamefont {Ritsch}}, \bibinfo {author} {\bibfnamefont
  {J.}~\bibnamefont {Schmiedmayer}}, \ and\ \bibinfo {author} {\bibfnamefont
  {J.}~\bibnamefont {Majer}},\ }\bibfield  {title} {\emph {\bibinfo {title}
  {Cavity qed with magnetically coupled collective spin states},\ }}\href
  {\doibase 10.1103/PhysRevLett.107.060502} {\bibfield  {journal} {\bibinfo
  {journal} {Phys. Rev. Lett.}\ }\textbf {\bibinfo {volume} {107}},\ \bibinfo
  {pages} {060502} (\bibinfo {year} {2011})}\BibitemShut {NoStop}%
\bibitem [{\citenamefont {Marcos}\ \emph {et~al.}(2010)\citenamefont {Marcos},
  \citenamefont {Wubs}, \citenamefont {Taylor}, \citenamefont {Aguado},
  \citenamefont {Lukin},\ and\ \citenamefont
  {S\o{}rensen}}]{Marcos2010coupling}%
  \BibitemOpen
  \bibfield  {author} {\bibinfo {author} {\bibfnamefont {D.}~\bibnamefont
  {Marcos}}, \bibinfo {author} {\bibfnamefont {M.}~\bibnamefont {Wubs}},
  \bibinfo {author} {\bibfnamefont {J.~M.}\ \bibnamefont {Taylor}}, \bibinfo
  {author} {\bibfnamefont {R.}~\bibnamefont {Aguado}}, \bibinfo {author}
  {\bibfnamefont {M.~D.}\ \bibnamefont {Lukin}}, \ and\ \bibinfo {author}
  {\bibfnamefont {A.~S.}\ \bibnamefont {S\o{}rensen}},\ }\bibfield  {title}
  {\emph {\bibinfo {title} {Coupling nitrogen-vacancy centers in diamond to
  superconducting flux qubits},\ }}\href {\doibase
  10.1103/PhysRevLett.105.210501} {\bibfield  {journal} {\bibinfo  {journal}
  {Phys. Rev. Lett.}\ }\textbf {\bibinfo {volume} {105}},\ \bibinfo {pages}
  {210501} (\bibinfo {year} {2010})}\BibitemShut {NoStop}%
\bibitem [{\citenamefont {Ranjan}\ \emph {et~al.}(2013)\citenamefont {Ranjan},
  \citenamefont {de~Lange}, \citenamefont {Schutjens}, \citenamefont
  {Debelhoir}, \citenamefont {Groen}, \citenamefont {Szombati}, \citenamefont
  {Thoen}, \citenamefont {Klapwijk}, \citenamefont {Hanson},\ and\
  \citenamefont {DiCarlo}}]{Ranjan2013probing}%
  \BibitemOpen
  \bibfield  {author} {\bibinfo {author} {\bibfnamefont {V.}~\bibnamefont
  {Ranjan}}, \bibinfo {author} {\bibfnamefont {G.}~\bibnamefont {de~Lange}},
  \bibinfo {author} {\bibfnamefont {R.}~\bibnamefont {Schutjens}}, \bibinfo
  {author} {\bibfnamefont {T.}~\bibnamefont {Debelhoir}}, \bibinfo {author}
  {\bibfnamefont {J.~P.}\ \bibnamefont {Groen}}, \bibinfo {author}
  {\bibfnamefont {D.}~\bibnamefont {Szombati}}, \bibinfo {author}
  {\bibfnamefont {D.~J.}\ \bibnamefont {Thoen}}, \bibinfo {author}
  {\bibfnamefont {T.~M.}\ \bibnamefont {Klapwijk}}, \bibinfo {author}
  {\bibfnamefont {R.}~\bibnamefont {Hanson}}, \ and\ \bibinfo {author}
  {\bibfnamefont {L.}~\bibnamefont {DiCarlo}},\ }\bibfield  {title} {\emph
  {\bibinfo {title} {Probing dynamics of an electron-spin ensemble via a
  superconducting resonator},\ }}\href {\doibase
  10.1103/PhysRevLett.110.067004} {\bibfield  {journal} {\bibinfo  {journal}
  {Phys. Rev. Lett.}\ }\textbf {\bibinfo {volume} {110}},\ \bibinfo {pages}
  {067004} (\bibinfo {year} {2013})}\BibitemShut {NoStop}%
\bibitem [{\citenamefont {Schuster}\ \emph {et~al.}(2010)\citenamefont
  {Schuster}, \citenamefont {Sears}, \citenamefont {Ginossar}, \citenamefont
  {DiCarlo}, \citenamefont {Frunzio}, \citenamefont {Morton}, \citenamefont
  {Wu}, \citenamefont {Briggs}, \citenamefont {Buckley}, \citenamefont
  {Awschalom},\ and\ \citenamefont {Schoelkopf}}]{Schuster2010high}%
  \BibitemOpen
  \bibfield  {author} {\bibinfo {author} {\bibfnamefont {D.~I.}\ \bibnamefont
  {Schuster}}, \bibinfo {author} {\bibfnamefont {A.~P.}\ \bibnamefont {Sears}},
  \bibinfo {author} {\bibfnamefont {E.}~\bibnamefont {Ginossar}}, \bibinfo
  {author} {\bibfnamefont {L.}~\bibnamefont {DiCarlo}}, \bibinfo {author}
  {\bibfnamefont {L.}~\bibnamefont {Frunzio}}, \bibinfo {author} {\bibfnamefont
  {J.~J.~L.}\ \bibnamefont {Morton}}, \bibinfo {author} {\bibfnamefont
  {H.}~\bibnamefont {Wu}}, \bibinfo {author} {\bibfnamefont {G.~A.~D.}\
  \bibnamefont {Briggs}}, \bibinfo {author} {\bibfnamefont {B.~B.}\
  \bibnamefont {Buckley}}, \bibinfo {author} {\bibfnamefont {D.~D.}\
  \bibnamefont {Awschalom}}, \ and\ \bibinfo {author} {\bibfnamefont {R.~J.}\
  \bibnamefont {Schoelkopf}},\ }\bibfield  {title} {\emph {\bibinfo {title}
  {High-cooperativity coupling of electron-spin ensembles to superconducting
  cavities},\ }}\href {\doibase 10.1103/PhysRevLett.105.140501} {\bibfield
  {journal} {\bibinfo  {journal} {Phys. Rev. Lett.}\ }\textbf {\bibinfo
  {volume} {105}},\ \bibinfo {pages} {140501} (\bibinfo {year}
  {2010})}\BibitemShut {NoStop}%
\bibitem [{\citenamefont {Probst}\ \emph {et~al.}(2013)\citenamefont {Probst},
  \citenamefont {Rotzinger}, \citenamefont {W\"unsch}, \citenamefont {Jung},
  \citenamefont {Jerger}, \citenamefont {Siegel}, \citenamefont {Ustinov},\
  and\ \citenamefont {Bushev}}]{Probst2013Anisotropic}%
  \BibitemOpen
  \bibfield  {author} {\bibinfo {author} {\bibfnamefont {S.}~\bibnamefont
  {Probst}}, \bibinfo {author} {\bibfnamefont {H.}~\bibnamefont {Rotzinger}},
  \bibinfo {author} {\bibfnamefont {S.}~\bibnamefont {W\"unsch}}, \bibinfo
  {author} {\bibfnamefont {P.}~\bibnamefont {Jung}}, \bibinfo {author}
  {\bibfnamefont {M.}~\bibnamefont {Jerger}}, \bibinfo {author} {\bibfnamefont
  {M.}~\bibnamefont {Siegel}}, \bibinfo {author} {\bibfnamefont {A.~V.}\
  \bibnamefont {Ustinov}}, \ and\ \bibinfo {author} {\bibfnamefont {P.~A.}\
  \bibnamefont {Bushev}},\ }\bibfield  {title} {\emph {\bibinfo {title}
  {Anisotropic rare-earth spin ensemble strongly coupled to a superconducting
  resonator},\ }}\href {\doibase 10.1103/PhysRevLett.110.157001} {\bibfield
  {journal} {\bibinfo  {journal} {Phys. Rev. Lett.}\ }\textbf {\bibinfo
  {volume} {110}},\ \bibinfo {pages} {157001} (\bibinfo {year}
  {2013})}\BibitemShut {NoStop}%
\bibitem [{\citenamefont {Tkal\ifmmode~\check{c}\else \v{c}\fi{}ec}\ \emph
  {et~al.}(2014)\citenamefont {Tkal\ifmmode~\check{c}\else \v{c}\fi{}ec},
  \citenamefont {Probst}, \citenamefont {Rieger}, \citenamefont {Rotzinger},
  \citenamefont {W\"unsch}, \citenamefont {Kukharchyk}, \citenamefont {Wieck},
  \citenamefont {Siegel}, \citenamefont {Ustinov},\ and\ \citenamefont
  {Bushev}}]{Tkal2014strong}%
  \BibitemOpen
  \bibfield  {author} {\bibinfo {author} {\bibfnamefont {A.}~\bibnamefont
  {Tkal\ifmmode~\check{c}\else \v{c}\fi{}ec}}, \bibinfo {author} {\bibfnamefont
  {S.}~\bibnamefont {Probst}}, \bibinfo {author} {\bibfnamefont
  {D.}~\bibnamefont {Rieger}}, \bibinfo {author} {\bibfnamefont
  {H.}~\bibnamefont {Rotzinger}}, \bibinfo {author} {\bibfnamefont
  {S.}~\bibnamefont {W\"unsch}}, \bibinfo {author} {\bibfnamefont
  {N.}~\bibnamefont {Kukharchyk}}, \bibinfo {author} {\bibfnamefont {A.~D.}\
  \bibnamefont {Wieck}}, \bibinfo {author} {\bibfnamefont {M.}~\bibnamefont
  {Siegel}}, \bibinfo {author} {\bibfnamefont {A.~V.}\ \bibnamefont {Ustinov}},
  \ and\ \bibinfo {author} {\bibfnamefont {P.}~\bibnamefont {Bushev}},\
  }\bibfield  {title} {\emph {\bibinfo {title} {Strong coupling of an
  {${\mathrm{Er}}^{3+}$}-doped {${\mathrm{YAlO}}_{3}$} crystal to a
  superconducting resonator},\ }}\href {\doibase 10.1103/PhysRevB.90.075112}
  {\bibfield  {journal} {\bibinfo  {journal} {Phys. Rev. B}\ }\textbf {\bibinfo
  {volume} {90}},\ \bibinfo {pages} {075112} (\bibinfo {year}
  {2014})}\BibitemShut {NoStop}%
\bibitem [{\citenamefont {Tabuchi}\ \emph {et~al.}(2014)\citenamefont
  {Tabuchi}, \citenamefont {Ishino}, \citenamefont {Ishikawa}, \citenamefont
  {Yamazaki}, \citenamefont {Usami},\ and\ \citenamefont
  {Nakamura}}]{Tabuchi2014Hybridizing}%
  \BibitemOpen
  \bibfield  {author} {\bibinfo {author} {\bibfnamefont {Y.}~\bibnamefont
  {Tabuchi}}, \bibinfo {author} {\bibfnamefont {S.}~\bibnamefont {Ishino}},
  \bibinfo {author} {\bibfnamefont {T.}~\bibnamefont {Ishikawa}}, \bibinfo
  {author} {\bibfnamefont {R.}~\bibnamefont {Yamazaki}}, \bibinfo {author}
  {\bibfnamefont {K.}~\bibnamefont {Usami}}, \ and\ \bibinfo {author}
  {\bibfnamefont {Y.}~\bibnamefont {Nakamura}},\ }\bibfield  {title} {\emph
  {\bibinfo {title} {Hybridizing ferromagnetic magnons and microwave photons in
  the quantum limit},\ }}\href {\doibase 10.1103/PhysRevLett.113.083603}
  {\bibfield  {journal} {\bibinfo  {journal} {Phys. Rev. Lett.}\ }\textbf
  {\bibinfo {volume} {113}},\ \bibinfo {pages} {083603} (\bibinfo {year}
  {2014})}\BibitemShut {NoStop}%
\bibitem [{\citenamefont {Huebl}\ \emph {et~al.}(2013)\citenamefont {Huebl},
  \citenamefont {Zollitsch}, \citenamefont {Lotze}, \citenamefont {Hocke},
  \citenamefont {Greifenstein}, \citenamefont {Marx}, \citenamefont {Gross},\
  and\ \citenamefont {Goennenwein}}]{Huebl2013High}%
  \BibitemOpen
  \bibfield  {author} {\bibinfo {author} {\bibfnamefont {H.}~\bibnamefont
  {Huebl}}, \bibinfo {author} {\bibfnamefont {C.~W.}\ \bibnamefont
  {Zollitsch}}, \bibinfo {author} {\bibfnamefont {J.}~\bibnamefont {Lotze}},
  \bibinfo {author} {\bibfnamefont {F.}~\bibnamefont {Hocke}}, \bibinfo
  {author} {\bibfnamefont {M.}~\bibnamefont {Greifenstein}}, \bibinfo {author}
  {\bibfnamefont {A.}~\bibnamefont {Marx}}, \bibinfo {author} {\bibfnamefont
  {R.}~\bibnamefont {Gross}}, \ and\ \bibinfo {author} {\bibfnamefont
  {S.~T.~B.}\ \bibnamefont {Goennenwein}},\ }\bibfield  {title} {\emph
  {\bibinfo {title} {High cooperativity in coupled microwave resonator
  ferrimagnetic insulator hybrids},\ }}\href {\doibase
  10.1103/PhysRevLett.111.127003} {\bibfield  {journal} {\bibinfo  {journal}
  {Phys. Rev. Lett.}\ }\textbf {\bibinfo {volume} {111}},\ \bibinfo {pages}
  {127003} (\bibinfo {year} {2013})}\BibitemShut {NoStop}%
\bibitem [{\citenamefont {Zhang}\ \emph {et~al.}(2014)\citenamefont {Zhang},
  \citenamefont {Zou}, \citenamefont {Jiang},\ and\ \citenamefont
  {Tang}}]{Zhang2014Strongly}%
  \BibitemOpen
  \bibfield  {author} {\bibinfo {author} {\bibfnamefont {X.}~\bibnamefont
  {Zhang}}, \bibinfo {author} {\bibfnamefont {C.-L.}\ \bibnamefont {Zou}},
  \bibinfo {author} {\bibfnamefont {L.}~\bibnamefont {Jiang}}, \ and\ \bibinfo
  {author} {\bibfnamefont {H.~X.}\ \bibnamefont {Tang}},\ }\bibfield  {title}
  {\emph {\bibinfo {title} {Strongly coupled magnons and cavity microwave
  photons},\ }}\href {\doibase 10.1103/PhysRevLett.113.156401} {\bibfield
  {journal} {\bibinfo  {journal} {Phys. Rev. Lett.}\ }\textbf {\bibinfo
  {volume} {113}},\ \bibinfo {pages} {156401} (\bibinfo {year}
  {2014})}\BibitemShut {NoStop}%
\bibitem [{\citenamefont {Goryachev}\ \emph {et~al.}(2014)\citenamefont
  {Goryachev}, \citenamefont {Farr}, \citenamefont {Creedon}, \citenamefont
  {Fan}, \citenamefont {Kostylev},\ and\ \citenamefont
  {Tobar}}]{Goryachev2014High}%
  \BibitemOpen
  \bibfield  {author} {\bibinfo {author} {\bibfnamefont {M.}~\bibnamefont
  {Goryachev}}, \bibinfo {author} {\bibfnamefont {W.~G.}\ \bibnamefont {Farr}},
  \bibinfo {author} {\bibfnamefont {D.~L.}\ \bibnamefont {Creedon}}, \bibinfo
  {author} {\bibfnamefont {Y.}~\bibnamefont {Fan}}, \bibinfo {author}
  {\bibfnamefont {M.}~\bibnamefont {Kostylev}}, \ and\ \bibinfo {author}
  {\bibfnamefont {M.~E.}\ \bibnamefont {Tobar}},\ }\bibfield  {title} {\emph
  {\bibinfo {title} {High-cooperativity cavity {QED} with magnons at microwave
  frequencies},\ }}\href {\doibase 10.1103/PhysRevApplied.2.054002} {\bibfield
  {journal} {\bibinfo  {journal} {Phys. Rev. Applied}\ }\textbf {\bibinfo
  {volume} {2}},\ \bibinfo {pages} {054002} (\bibinfo {year}
  {2014})}\BibitemShut {NoStop}%
\bibitem [{\citenamefont {Bai}\ \emph {et~al.}(2015)\citenamefont {Bai},
  \citenamefont {Harder}, \citenamefont {Chen}, \citenamefont {Fan},
  \citenamefont {Xiao},\ and\ \citenamefont {Hu}}]{Bai2015Spin}%
  \BibitemOpen
  \bibfield  {author} {\bibinfo {author} {\bibfnamefont {L.}~\bibnamefont
  {Bai}}, \bibinfo {author} {\bibfnamefont {M.}~\bibnamefont {Harder}},
  \bibinfo {author} {\bibfnamefont {Y.~P.}\ \bibnamefont {Chen}}, \bibinfo
  {author} {\bibfnamefont {X.}~\bibnamefont {Fan}}, \bibinfo {author}
  {\bibfnamefont {J.~Q.}\ \bibnamefont {Xiao}}, \ and\ \bibinfo {author}
  {\bibfnamefont {C.-M.}\ \bibnamefont {Hu}},\ }\bibfield  {title} {\emph
  {\bibinfo {title} {Spin pumping in electrodynamically coupled magnon-photon
  systems},\ }}\href {\doibase 10.1103/PhysRevLett.114.227201} {\bibfield
  {journal} {\bibinfo  {journal} {Phys. Rev. Lett.}\ }\textbf {\bibinfo
  {volume} {114}},\ \bibinfo {pages} {227201} (\bibinfo {year}
  {2015})}\BibitemShut {NoStop}%
\bibitem [{\citenamefont {Soykal}\ and\ \citenamefont
  {Flatt\'e}(2010)}]{Soykal2010Strong}%
  \BibitemOpen
  \bibfield  {author} {\bibinfo {author} {\bibfnamefont {O.~O.}\ \bibnamefont
  {Soykal}}\ and\ \bibinfo {author} {\bibfnamefont {M.~E.}\ \bibnamefont
  {Flatt\'e}},\ }\bibfield  {title} {\emph {\bibinfo {title} {Strong field
  interactions between a nanomagnet and a photonic cavity},\ }}\href {\doibase
  10.1103/PhysRevLett.104.077202} {\bibfield  {journal} {\bibinfo  {journal}
  {Phys. Rev. Lett.}\ }\textbf {\bibinfo {volume} {104}},\ \bibinfo {pages}
  {077202} (\bibinfo {year} {2010})}\BibitemShut {NoStop}%
\bibitem [{\citenamefont {Bourhill}\ \emph {et~al.}(2016)\citenamefont
  {Bourhill}, \citenamefont {Kostylev}, \citenamefont {Goryachev},
  \citenamefont {Creedon},\ and\ \citenamefont
  {Tobar}}]{Bourhill2016Ultrahigh}%
  \BibitemOpen
  \bibfield  {author} {\bibinfo {author} {\bibfnamefont {J.}~\bibnamefont
  {Bourhill}}, \bibinfo {author} {\bibfnamefont {N.}~\bibnamefont {Kostylev}},
  \bibinfo {author} {\bibfnamefont {M.}~\bibnamefont {Goryachev}}, \bibinfo
  {author} {\bibfnamefont {D.~L.}\ \bibnamefont {Creedon}}, \ and\ \bibinfo
  {author} {\bibfnamefont {M.~E.}\ \bibnamefont {Tobar}},\ }\bibfield  {title}
  {\emph {\bibinfo {title} {Ultrahigh cooperativity interactions between
  magnons and resonant photons in a {YIG} sphere},\ }}\href {\doibase
  10.1103/PhysRevB.93.144420} {\bibfield  {journal} {\bibinfo  {journal} {Phys.
  Rev. B}\ }\textbf {\bibinfo {volume} {93}},\ \bibinfo {pages} {144420}
  (\bibinfo {year} {2016})}\BibitemShut {NoStop}%
\bibitem [{\citenamefont {Z.~Rameshti}\ \emph {et~al.}(2015)\citenamefont
  {Z.~Rameshti}, \citenamefont {Cao},\ and\ \citenamefont
  {Bauer}}]{Zare2015Magnetic}%
  \BibitemOpen
  \bibfield  {author} {\bibinfo {author} {\bibfnamefont {B.}~\bibnamefont
  {Z.~Rameshti}}, \bibinfo {author} {\bibfnamefont {Y.}~\bibnamefont {Cao}}, \
  and\ \bibinfo {author} {\bibfnamefont {G.~E.~W.}\ \bibnamefont {Bauer}},\
  }\bibfield  {title} {\emph {\bibinfo {title} {Magnetic spheres in microwave
  cavities},\ }}\href {\doibase 10.1103/PhysRevB.91.214430} {\bibfield
  {journal} {\bibinfo  {journal} {Phys. Rev. B}\ }\textbf {\bibinfo {volume}
  {91}},\ \bibinfo {pages} {214430} (\bibinfo {year} {2015})}\BibitemShut
  {NoStop}%
\bibitem [{\citenamefont {V.~Kusminskiy}\ \emph {et~al.}(2016)\citenamefont
  {V.~Kusminskiy}, \citenamefont {Tang},\ and\ \citenamefont
  {Marquardt}}]{Viola2016Coupled}%
  \BibitemOpen
  \bibfield  {author} {\bibinfo {author} {\bibfnamefont {S.}~\bibnamefont
  {V.~Kusminskiy}}, \bibinfo {author} {\bibfnamefont {H.~X.}\ \bibnamefont
  {Tang}}, \ and\ \bibinfo {author} {\bibfnamefont {F.}~\bibnamefont
  {Marquardt}},\ }\bibfield  {title} {\emph {\bibinfo {title} {Coupled
  spin-light dynamics in cavity optomagnonics},\ }}\href {\doibase
  10.1103/PhysRevA.94.033821} {\bibfield  {journal} {\bibinfo  {journal} {Phys.
  Rev. A}\ }\textbf {\bibinfo {volume} {94}},\ \bibinfo {pages} {033821}
  (\bibinfo {year} {2016})}\BibitemShut {NoStop}%
\bibitem [{\citenamefont {Zhang}\ \emph {et~al.}(2016)\citenamefont {Zhang},
  \citenamefont {Zhu}, \citenamefont {Zou},\ and\ \citenamefont
  {Tang}}]{Zhang2016Opto}%
  \BibitemOpen
  \bibfield  {author} {\bibinfo {author} {\bibfnamefont {X.}~\bibnamefont
  {Zhang}}, \bibinfo {author} {\bibfnamefont {N.}~\bibnamefont {Zhu}}, \bibinfo
  {author} {\bibfnamefont {C.-L.}\ \bibnamefont {Zou}}, \ and\ \bibinfo
  {author} {\bibfnamefont {H.~X.}\ \bibnamefont {Tang}},\ }\bibfield  {title}
  {\emph {\bibinfo {title} {Optomagnonic whispering gallery microresonators},\
  }}\href {\doibase 10.1103/PhysRevLett.117.123605} {\bibfield  {journal}
  {\bibinfo  {journal} {Phys. Rev. Lett.}\ }\textbf {\bibinfo {volume} {117}},\
  \bibinfo {pages} {123605} (\bibinfo {year} {2016})}\BibitemShut {NoStop}%
\bibitem [{\citenamefont {Osada}\ \emph {et~al.}(2016)\citenamefont {Osada},
  \citenamefont {Hisatomi}, \citenamefont {Noguchi}, \citenamefont {Tabuchi},
  \citenamefont {Yamazaki}, \citenamefont {Usami}, \citenamefont {Sadgrove},
  \citenamefont {Yalla}, \citenamefont {Nomura},\ and\ \citenamefont
  {Nakamura}}]{Osada2016Cavity}%
  \BibitemOpen
  \bibfield  {author} {\bibinfo {author} {\bibfnamefont {A.}~\bibnamefont
  {Osada}}, \bibinfo {author} {\bibfnamefont {R.}~\bibnamefont {Hisatomi}},
  \bibinfo {author} {\bibfnamefont {A.}~\bibnamefont {Noguchi}}, \bibinfo
  {author} {\bibfnamefont {Y.}~\bibnamefont {Tabuchi}}, \bibinfo {author}
  {\bibfnamefont {R.}~\bibnamefont {Yamazaki}}, \bibinfo {author}
  {\bibfnamefont {K.}~\bibnamefont {Usami}}, \bibinfo {author} {\bibfnamefont
  {M.}~\bibnamefont {Sadgrove}}, \bibinfo {author} {\bibfnamefont
  {R.}~\bibnamefont {Yalla}}, \bibinfo {author} {\bibfnamefont
  {M.}~\bibnamefont {Nomura}}, \ and\ \bibinfo {author} {\bibfnamefont
  {Y.}~\bibnamefont {Nakamura}},\ }\bibfield  {title} {\emph {\bibinfo {title}
  {Cavity optomagnonics with spin-orbit coupled photons},\ }}\href {\doibase
  10.1103/PhysRevLett.116.223601} {\bibfield  {journal} {\bibinfo  {journal}
  {Phys. Rev. Lett.}\ }\textbf {\bibinfo {volume} {116}},\ \bibinfo {pages}
  {223601} (\bibinfo {year} {2016})}\BibitemShut {NoStop}%
\bibitem [{\citenamefont {Haigh}\ \emph {et~al.}(2016)\citenamefont {Haigh},
  \citenamefont {Nunnenkamp}, \citenamefont {Ramsay},\ and\ \citenamefont
  {Ferguson}}]{Haigh2016Triple}%
  \BibitemOpen
  \bibfield  {author} {\bibinfo {author} {\bibfnamefont {J.~A.}\ \bibnamefont
  {Haigh}}, \bibinfo {author} {\bibfnamefont {A.}~\bibnamefont {Nunnenkamp}},
  \bibinfo {author} {\bibfnamefont {A.~J.}\ \bibnamefont {Ramsay}}, \ and\
  \bibinfo {author} {\bibfnamefont {A.~J.}\ \bibnamefont {Ferguson}},\
  }\bibfield  {title} {\emph {\bibinfo {title} {Triple-resonant {B}rillouin
  light scattering in magneto-optical cavities},\ }}\href {\doibase
  10.1103/PhysRevLett.117.133602} {\bibfield  {journal} {\bibinfo  {journal}
  {Phys. Rev. Lett.}\ }\textbf {\bibinfo {volume} {117}},\ \bibinfo {pages}
  {133602} (\bibinfo {year} {2016})}\BibitemShut {NoStop}%
\bibitem [{\citenamefont {Liu}\ \emph {et~al.}(2019)\citenamefont {Liu},
  \citenamefont {Xiong},\ and\ \citenamefont {Wu}}]{Liu2019Magnon}%
  \BibitemOpen
  \bibfield  {author} {\bibinfo {author} {\bibfnamefont {Z.-X.}\ \bibnamefont
  {Liu}}, \bibinfo {author} {\bibfnamefont {H.}~\bibnamefont {Xiong}}, \ and\
  \bibinfo {author} {\bibfnamefont {Y.}~\bibnamefont {Wu}},\ }\bibfield
  {title} {\emph {\bibinfo {title} {Magnon blockade in a hybrid
  ferromagnet-superconductor quantum system},\ }}\href {\doibase
  10.1103/PhysRevB.100.134421} {\bibfield  {journal} {\bibinfo  {journal}
  {Phys. Rev. B}\ }\textbf {\bibinfo {volume} {100}},\ \bibinfo {pages}
  {134421} (\bibinfo {year} {2019})}\BibitemShut {NoStop}%
\bibitem [{\citenamefont {Xie}\ \emph {et~al.}(2020)\citenamefont {Xie},
  \citenamefont {Ma},\ and\ \citenamefont {Li}}]{Xie2020Quantum}%
  \BibitemOpen
  \bibfield  {author} {\bibinfo {author} {\bibfnamefont {J.-k.}\ \bibnamefont
  {Xie}}, \bibinfo {author} {\bibfnamefont {S.-l.}\ \bibnamefont {Ma}}, \ and\
  \bibinfo {author} {\bibfnamefont {F.-l.}\ \bibnamefont {Li}},\ }\bibfield
  {title} {\emph {\bibinfo {title} {Quantum-interference-enhanced magnon
  blockade in an yttrium-iron-garnet sphere coupled to superconducting
  circuits},\ }}\href {\doibase 10.1103/PhysRevA.101.042331} {\bibfield
  {journal} {\bibinfo  {journal} {Phys. Rev. A}\ }\textbf {\bibinfo {volume}
  {101}},\ \bibinfo {pages} {042331} (\bibinfo {year} {2020})}\BibitemShut
  {NoStop}%
\bibitem [{\citenamefont {Tabuchi}\ \emph {et~al.}(2015)\citenamefont
  {Tabuchi}, \citenamefont {Ishino}, \citenamefont {Noguchi}, \citenamefont
  {Ishikawa}, \citenamefont {Yamazaki}, \citenamefont {Usami},\ and\
  \citenamefont {Nakamura}}]{Yutaka2015Coherent}%
  \BibitemOpen
  \bibfield  {author} {\bibinfo {author} {\bibfnamefont {Y.}~\bibnamefont
  {Tabuchi}}, \bibinfo {author} {\bibfnamefont {S.}~\bibnamefont {Ishino}},
  \bibinfo {author} {\bibfnamefont {A.}~\bibnamefont {Noguchi}}, \bibinfo
  {author} {\bibfnamefont {T.}~\bibnamefont {Ishikawa}}, \bibinfo {author}
  {\bibfnamefont {R.}~\bibnamefont {Yamazaki}}, \bibinfo {author}
  {\bibfnamefont {K.}~\bibnamefont {Usami}}, \ and\ \bibinfo {author}
  {\bibfnamefont {Y.}~\bibnamefont {Nakamura}},\ }\bibfield  {title} {\emph
  {\bibinfo {title} {Coherent coupling between a ferromagnetic magnon and a
  superconducting qubit},\ }}\href {\doibase 10.1126/science.aaa3693}
  {\bibfield  {journal} {\bibinfo  {journal} {Science}\ }\textbf {\bibinfo
  {volume} {349}},\ \bibinfo {pages} {405} (\bibinfo {year}
  {2015})}\BibitemShut {NoStop}%
\bibitem [{\citenamefont {Lachance-Quirion}\ \emph {et~al.}(2017)\citenamefont
  {Lachance-Quirion}, \citenamefont {Tabuchi}, \citenamefont {Ishino},
  \citenamefont {Noguchi}, \citenamefont {Ishikawa}, \citenamefont {Yamazaki},\
  and\ \citenamefont {Nakamura}}]{Dany2017Resolving}%
  \BibitemOpen
  \bibfield  {author} {\bibinfo {author} {\bibfnamefont {D.}~\bibnamefont
  {Lachance-Quirion}}, \bibinfo {author} {\bibfnamefont {Y.}~\bibnamefont
  {Tabuchi}}, \bibinfo {author} {\bibfnamefont {S.}~\bibnamefont {Ishino}},
  \bibinfo {author} {\bibfnamefont {A.}~\bibnamefont {Noguchi}}, \bibinfo
  {author} {\bibfnamefont {T.}~\bibnamefont {Ishikawa}}, \bibinfo {author}
  {\bibfnamefont {R.}~\bibnamefont {Yamazaki}}, \ and\ \bibinfo {author}
  {\bibfnamefont {Y.}~\bibnamefont {Nakamura}},\ }\bibfield  {title} {\emph
  {\bibinfo {title} {Resolving quanta of collective spin excitations in a
  millimeter-sized ferromagnet},\ }}\href {\doibase 10.1126/sciadv.1603150}
  {\bibfield  {journal} {\bibinfo  {journal} {Sci. Adv.}\ }\textbf {\bibinfo
  {volume} {3}},\ \bibinfo {pages} {e1603150} (\bibinfo {year}
  {2017})}\BibitemShut {NoStop}%
\bibitem [{\citenamefont {Wolski}\ \emph {et~al.}(2020)\citenamefont {Wolski},
  \citenamefont {Lachance-Quirion}, \citenamefont {Tabuchi}, \citenamefont
  {Kono}, \citenamefont {Noguchi}, \citenamefont {Usami},\ and\ \citenamefont
  {Nakamura}}]{Wolski2020Dissipation}%
  \BibitemOpen
  \bibfield  {author} {\bibinfo {author} {\bibfnamefont {S.~P.}\ \bibnamefont
  {Wolski}}, \bibinfo {author} {\bibfnamefont {D.}~\bibnamefont
  {Lachance-Quirion}}, \bibinfo {author} {\bibfnamefont {Y.}~\bibnamefont
  {Tabuchi}}, \bibinfo {author} {\bibfnamefont {S.}~\bibnamefont {Kono}},
  \bibinfo {author} {\bibfnamefont {A.}~\bibnamefont {Noguchi}}, \bibinfo
  {author} {\bibfnamefont {K.}~\bibnamefont {Usami}}, \ and\ \bibinfo {author}
  {\bibfnamefont {Y.}~\bibnamefont {Nakamura}},\ }\bibfield  {title} {\emph
  {\bibinfo {title} {Dissipation-based quantum sensing of magnons with a
  superconducting qubit},\ }}\href {\doibase 10.1103/PhysRevLett.125.117701}
  {\bibfield  {journal} {\bibinfo  {journal} {Phys. Rev. Lett.}\ }\textbf
  {\bibinfo {volume} {125}},\ \bibinfo {pages} {117701} (\bibinfo {year}
  {2020})}\BibitemShut {NoStop}%
\bibitem [{\citenamefont {Lachance-Quirion}\ \emph {et~al.}(2020)\citenamefont
  {Lachance-Quirion}, \citenamefont {Wolski}, \citenamefont {Tabuchi},
  \citenamefont {Kono}, \citenamefont {Usami},\ and\ \citenamefont
  {Nakamura}}]{Dany2020Entanglement}%
  \BibitemOpen
  \bibfield  {author} {\bibinfo {author} {\bibfnamefont {D.}~\bibnamefont
  {Lachance-Quirion}}, \bibinfo {author} {\bibfnamefont {S.~P.}\ \bibnamefont
  {Wolski}}, \bibinfo {author} {\bibfnamefont {Y.}~\bibnamefont {Tabuchi}},
  \bibinfo {author} {\bibfnamefont {S.}~\bibnamefont {Kono}}, \bibinfo {author}
  {\bibfnamefont {K.}~\bibnamefont {Usami}}, \ and\ \bibinfo {author}
  {\bibfnamefont {Y.}~\bibnamefont {Nakamura}},\ }\bibfield  {title} {\emph
  {\bibinfo {title} {Entanglement-based single-shot detection of a single
  magnon with a superconducting qubit},\ }}\href {\doibase
  10.1126/science.aaz9236} {\bibfield  {journal} {\bibinfo  {journal}
  {Science}\ }\textbf {\bibinfo {volume} {367}},\ \bibinfo {pages} {425}
  (\bibinfo {year} {2020})}\BibitemShut {NoStop}%
\bibitem [{\citenamefont {Kounalakis}\ \emph {et~al.}(2022)\citenamefont
  {Kounalakis}, \citenamefont {Bauer},\ and\ \citenamefont
  {Blanter}}]{Kounalakis2022Analog}%
  \BibitemOpen
  \bibfield  {author} {\bibinfo {author} {\bibfnamefont {M.}~\bibnamefont
  {Kounalakis}}, \bibinfo {author} {\bibfnamefont {G.~E.~W.}\ \bibnamefont
  {Bauer}}, \ and\ \bibinfo {author} {\bibfnamefont {Y.~M.}\ \bibnamefont
  {Blanter}},\ }\bibfield  {title} {\emph {\bibinfo {title} {Analog quantum
  control of magnonic cat states on a chip by a superconducting qubit},\
  }}\href {\doibase 10.1103/PhysRevLett.129.037205} {\bibfield  {journal}
  {\bibinfo  {journal} {Phys. Rev. Lett.}\ }\textbf {\bibinfo {volume} {129}},\
  \bibinfo {pages} {037205} (\bibinfo {year} {2022})}\BibitemShut {NoStop}%
\bibitem [{\citenamefont {Wang}\ \emph {et~al.}(2018)\citenamefont {Wang},
  \citenamefont {Zhang}, \citenamefont {Zhang}, \citenamefont {Li},
  \citenamefont {Hu},\ and\ \citenamefont {You}}]{wang2018bistability}%
  \BibitemOpen
  \bibfield  {author} {\bibinfo {author} {\bibfnamefont {Y.-P.}\ \bibnamefont
  {Wang}}, \bibinfo {author} {\bibfnamefont {G.-Q.}\ \bibnamefont {Zhang}},
  \bibinfo {author} {\bibfnamefont {D.}~\bibnamefont {Zhang}}, \bibinfo
  {author} {\bibfnamefont {T.-F.}\ \bibnamefont {Li}}, \bibinfo {author}
  {\bibfnamefont {C.-M.}\ \bibnamefont {Hu}}, \ and\ \bibinfo {author}
  {\bibfnamefont {J.~Q.}\ \bibnamefont {You}},\ }\bibfield  {title} {\emph
  {\bibinfo {title} {Bistability of cavity magnon polaritons},\ }}\href
  {\doibase 10.1103/PhysRevLett.120.057202} {\bibfield  {journal} {\bibinfo
  {journal} {Phys. Rev. Lett.}\ }\textbf {\bibinfo {volume} {120}},\ \bibinfo
  {pages} {057202} (\bibinfo {year} {2018})}\BibitemShut {NoStop}%
\bibitem [{\citenamefont {Bai}\ \emph {et~al.}(2017)\citenamefont {Bai},
  \citenamefont {Harder}, \citenamefont {Hyde}, \citenamefont {Zhang},
  \citenamefont {Hu}, \citenamefont {Chen},\ and\ \citenamefont
  {Xiao}}]{Bai2017Cavity}%
  \BibitemOpen
  \bibfield  {author} {\bibinfo {author} {\bibfnamefont {L.}~\bibnamefont
  {Bai}}, \bibinfo {author} {\bibfnamefont {M.}~\bibnamefont {Harder}},
  \bibinfo {author} {\bibfnamefont {P.}~\bibnamefont {Hyde}}, \bibinfo {author}
  {\bibfnamefont {Z.}~\bibnamefont {Zhang}}, \bibinfo {author} {\bibfnamefont
  {C.-M.}\ \bibnamefont {Hu}}, \bibinfo {author} {\bibfnamefont {Y.~P.}\
  \bibnamefont {Chen}}, \ and\ \bibinfo {author} {\bibfnamefont {J.~Q.}\
  \bibnamefont {Xiao}},\ }\bibfield  {title} {\emph {\bibinfo {title} {Cavity
  mediated manipulation of distant spin currents using a
  cavity-magnon-polariton},\ }}\href {\doibase 10.1103/PhysRevLett.118.217201}
  {\bibfield  {journal} {\bibinfo  {journal} {Phys. Rev. Lett.}\ }\textbf
  {\bibinfo {volume} {118}},\ \bibinfo {pages} {217201} (\bibinfo {year}
  {2017})}\BibitemShut {NoStop}%
\bibitem [{\citenamefont {Sharma}\ \emph {et~al.}(2018)\citenamefont {Sharma},
  \citenamefont {Blanter},\ and\ \citenamefont {Bauer}}]{Sharma2018Optical}%
  \BibitemOpen
  \bibfield  {author} {\bibinfo {author} {\bibfnamefont {S.}~\bibnamefont
  {Sharma}}, \bibinfo {author} {\bibfnamefont {Y.~M.}\ \bibnamefont {Blanter}},
  \ and\ \bibinfo {author} {\bibfnamefont {G.~E.~W.}\ \bibnamefont {Bauer}},\
  }\bibfield  {title} {\emph {\bibinfo {title} {Optical cooling of magnons},\
  }}\href {\doibase 10.1103/PhysRevLett.121.087205} {\bibfield  {journal}
  {\bibinfo  {journal} {Phys. Rev. Lett.}\ }\textbf {\bibinfo {volume} {121}},\
  \bibinfo {pages} {087205} (\bibinfo {year} {2018})}\BibitemShut {NoStop}%
\bibitem [{\citenamefont {Zhang}\ \emph {et~al.}(2015)\citenamefont {Zhang},
  \citenamefont {Zou}, \citenamefont {Zhu}, \citenamefont {Marquardt},
  \citenamefont {Jiang},\ and\ \citenamefont {Tang}}]{Zhang2015magnon}%
  \BibitemOpen
  \bibfield  {author} {\bibinfo {author} {\bibfnamefont {X.-F.}\ \bibnamefont
  {Zhang}}, \bibinfo {author} {\bibfnamefont {C.-L.}\ \bibnamefont {Zou}},
  \bibinfo {author} {\bibfnamefont {N.}~\bibnamefont {Zhu}}, \bibinfo {author}
  {\bibfnamefont {F.}~\bibnamefont {Marquardt}}, \bibinfo {author}
  {\bibfnamefont {L.}~\bibnamefont {Jiang}}, \ and\ \bibinfo {author}
  {\bibfnamefont {H.-X.}\ \bibnamefont {Tang}},\ }\bibfield  {title} {\emph
  {\bibinfo {title} {Magnon dark modes and gradient memory},\ }}\href {\doibase
  10.1038/ncomms9914} {\bibfield  {journal} {\bibinfo  {journal} {Nat.
  Commun.}\ }\textbf {\bibinfo {volume} {6}},\ \bibinfo {pages} {8914}
  (\bibinfo {year} {2015})}\BibitemShut {NoStop}%
\bibitem [{\citenamefont {Yuan}\ \emph
  {et~al.}(2020{\natexlab{a}})\citenamefont {Yuan}, \citenamefont {Zheng},
  \citenamefont {Ficek}, \citenamefont {He},\ and\ \citenamefont
  {Yung}}]{Yuan2020enhancement}%
  \BibitemOpen
  \bibfield  {author} {\bibinfo {author} {\bibfnamefont {H.~Y.}\ \bibnamefont
  {Yuan}}, \bibinfo {author} {\bibfnamefont {S.}~\bibnamefont {Zheng}},
  \bibinfo {author} {\bibfnamefont {Z.}~\bibnamefont {Ficek}}, \bibinfo
  {author} {\bibfnamefont {Q.~Y.}\ \bibnamefont {He}}, \ and\ \bibinfo {author}
  {\bibfnamefont {M.-H.}\ \bibnamefont {Yung}},\ }\bibfield  {title} {\emph
  {\bibinfo {title} {Enhancement of magnon-magnon entanglement inside a
  cavity},\ }}\href {\doibase 10.1103/PhysRevB.101.014419} {\bibfield
  {journal} {\bibinfo  {journal} {Phys. Rev. B}\ }\textbf {\bibinfo {volume}
  {101}},\ \bibinfo {pages} {014419} (\bibinfo {year}
  {2020}{\natexlab{a}})}\BibitemShut {NoStop}%
\bibitem [{\citenamefont {A.~Mousolou}\ \emph {et~al.}(2021)\citenamefont
  {A.~Mousolou}, \citenamefont {Liu}, \citenamefont {Bergman}, \citenamefont
  {Delin}, \citenamefont {Eriksson}, \citenamefont {Pereiro}, \citenamefont
  {Thonig},\ and\ \citenamefont {Sj\"oqvist}}]{Azimi2021Magnon}%
  \BibitemOpen
  \bibfield  {author} {\bibinfo {author} {\bibfnamefont {V.}~\bibnamefont
  {A.~Mousolou}}, \bibinfo {author} {\bibfnamefont {Y.}~\bibnamefont {Liu}},
  \bibinfo {author} {\bibfnamefont {A.}~\bibnamefont {Bergman}}, \bibinfo
  {author} {\bibfnamefont {A.}~\bibnamefont {Delin}}, \bibinfo {author}
  {\bibfnamefont {O.}~\bibnamefont {Eriksson}}, \bibinfo {author}
  {\bibfnamefont {M.}~\bibnamefont {Pereiro}}, \bibinfo {author} {\bibfnamefont
  {D.}~\bibnamefont {Thonig}}, \ and\ \bibinfo {author} {\bibfnamefont
  {E.}~\bibnamefont {Sj\"oqvist}},\ }\bibfield  {title} {\emph {\bibinfo
  {title} {Magnon-magnon entanglement and its quantification via a microwave
  cavity},\ }}\href {\doibase 10.1103/PhysRevB.104.224302} {\bibfield
  {journal} {\bibinfo  {journal} {Phys. Rev. B}\ }\textbf {\bibinfo {volume}
  {104}},\ \bibinfo {pages} {224302} (\bibinfo {year} {2021})}\BibitemShut
  {NoStop}%
\bibitem [{\citenamefont {Ren}\ \emph {et~al.}(2022)\citenamefont {Ren},
  \citenamefont {Xie}, \citenamefont {Li}, \citenamefont {Ma},\ and\
  \citenamefont {Li}}]{Ren2022Long}%
  \BibitemOpen
  \bibfield  {author} {\bibinfo {author} {\bibfnamefont {Y.-l.}\ \bibnamefont
  {Ren}}, \bibinfo {author} {\bibfnamefont {J.-k.}\ \bibnamefont {Xie}},
  \bibinfo {author} {\bibfnamefont {X.-k.}\ \bibnamefont {Li}}, \bibinfo
  {author} {\bibfnamefont {S.-l.}\ \bibnamefont {Ma}}, \ and\ \bibinfo {author}
  {\bibfnamefont {F.-l.}\ \bibnamefont {Li}},\ }\bibfield  {title} {\emph
  {\bibinfo {title} {Long-range generation of a magnon-magnon entangled
  state},\ }}\href {\doibase 10.1103/PhysRevB.105.094422} {\bibfield  {journal}
  {\bibinfo  {journal} {Phys. Rev. B}\ }\textbf {\bibinfo {volume} {105}},\
  \bibinfo {pages} {094422} (\bibinfo {year} {2022})}\BibitemShut {NoStop}%
\bibitem [{\citenamefont {Li}\ \emph {et~al.}(2018)\citenamefont {Li},
  \citenamefont {Zhu},\ and\ \citenamefont {Agarwal}}]{Li2018Magnon}%
  \BibitemOpen
  \bibfield  {author} {\bibinfo {author} {\bibfnamefont {J.}~\bibnamefont
  {Li}}, \bibinfo {author} {\bibfnamefont {S.-Y.}\ \bibnamefont {Zhu}}, \ and\
  \bibinfo {author} {\bibfnamefont {G.~S.}\ \bibnamefont {Agarwal}},\
  }\bibfield  {title} {\emph {\bibinfo {title} {Magnon-photon-phonon
  entanglement in cavity magnomechanics},\ }}\href {\doibase
  10.1103/PhysRevLett.121.203601} {\bibfield  {journal} {\bibinfo  {journal}
  {Phys. Rev. Lett.}\ }\textbf {\bibinfo {volume} {121}},\ \bibinfo {pages}
  {203601} (\bibinfo {year} {2018})}\BibitemShut {NoStop}%
\bibitem [{\citenamefont {Yuan}\ \emph
  {et~al.}(2020{\natexlab{b}})\citenamefont {Yuan}, \citenamefont {Yan},
  \citenamefont {Zheng}, \citenamefont {He}, \citenamefont {Xia},\ and\
  \citenamefont {Yung}}]{Yuan2020Steady}%
  \BibitemOpen
  \bibfield  {author} {\bibinfo {author} {\bibfnamefont {H.~Y.}\ \bibnamefont
  {Yuan}}, \bibinfo {author} {\bibfnamefont {P.}~\bibnamefont {Yan}}, \bibinfo
  {author} {\bibfnamefont {S.}~\bibnamefont {Zheng}}, \bibinfo {author}
  {\bibfnamefont {Q.~Y.}\ \bibnamefont {He}}, \bibinfo {author} {\bibfnamefont
  {K.}~\bibnamefont {Xia}}, \ and\ \bibinfo {author} {\bibfnamefont {M.-H.}\
  \bibnamefont {Yung}},\ }\bibfield  {title} {\emph {\bibinfo {title} {Steady
  {B}ell state generation via magnon-photon coupling},\ }}\href {\doibase
  10.1103/PhysRevLett.124.053602} {\bibfield  {journal} {\bibinfo  {journal}
  {Phys. Rev. Lett.}\ }\textbf {\bibinfo {volume} {124}},\ \bibinfo {pages}
  {053602} (\bibinfo {year} {2020}{\natexlab{b}})}\BibitemShut {NoStop}%
\bibitem [{\citenamefont {Xu}\ \emph {et~al.}(2023)\citenamefont {Xu},
  \citenamefont {Gu}, \citenamefont {Li}, \citenamefont {Weng}, \citenamefont
  {Wang}, \citenamefont {Li}, \citenamefont {Wang}, \citenamefont {Zhu},\ and\
  \citenamefont {You}}]{Xu2022Quantum}%
  \BibitemOpen
  \bibfield  {author} {\bibinfo {author} {\bibfnamefont {D.}~\bibnamefont
  {Xu}}, \bibinfo {author} {\bibfnamefont {X.-K.}\ \bibnamefont {Gu}}, \bibinfo
  {author} {\bibfnamefont {H.-K.}\ \bibnamefont {Li}}, \bibinfo {author}
  {\bibfnamefont {Y.-C.}\ \bibnamefont {Weng}}, \bibinfo {author}
  {\bibfnamefont {Y.-P.}\ \bibnamefont {Wang}}, \bibinfo {author}
  {\bibfnamefont {J.}~\bibnamefont {Li}}, \bibinfo {author} {\bibfnamefont
  {H.}~\bibnamefont {Wang}}, \bibinfo {author} {\bibfnamefont {S.-Y.}\
  \bibnamefont {Zhu}}, \ and\ \bibinfo {author} {\bibfnamefont {J.~Q.}\
  \bibnamefont {You}},\ }\bibfield  {title} {\emph {\bibinfo {title} {Quantum
  control of a single magnon in a macroscopic spin system},\ }}\href {\doibase
  10.1103/PhysRevLett.130.193603} {\bibfield  {journal} {\bibinfo  {journal}
  {Phys. Rev. Lett.}\ }\textbf {\bibinfo {volume} {130}},\ \bibinfo {pages}
  {193603} (\bibinfo {year} {2023})}\BibitemShut {NoStop}%
\bibitem [{\citenamefont {Zheng}\ \emph {et~al.}(2011)\citenamefont {Zheng},
  \citenamefont {Gauthier},\ and\ \citenamefont {Baranger}}]{Zheng2011Cavity}%
  \BibitemOpen
  \bibfield  {author} {\bibinfo {author} {\bibfnamefont {H.}~\bibnamefont
  {Zheng}}, \bibinfo {author} {\bibfnamefont {D.~J.}\ \bibnamefont {Gauthier}},
  \ and\ \bibinfo {author} {\bibfnamefont {H.~U.}\ \bibnamefont {Baranger}},\
  }\bibfield  {title} {\emph {\bibinfo {title} {Cavity-free photon blockade
  induced by many-body bound states},\ }}\href {\doibase
  10.1103/PhysRevLett.107.223601} {\bibfield  {journal} {\bibinfo  {journal}
  {Phys. Rev. Lett.}\ }\textbf {\bibinfo {volume} {107}},\ \bibinfo {pages}
  {223601} (\bibinfo {year} {2011})}\BibitemShut {NoStop}%
\bibitem [{\citenamefont {Huang}\ \emph {et~al.}(2013)\citenamefont {Huang},
  \citenamefont {Liao},\ and\ \citenamefont {Sun}}]{Huang2013photon}%
  \BibitemOpen
  \bibfield  {author} {\bibinfo {author} {\bibfnamefont {J.-F.}\ \bibnamefont
  {Huang}}, \bibinfo {author} {\bibfnamefont {J.-Q.}\ \bibnamefont {Liao}}, \
  and\ \bibinfo {author} {\bibfnamefont {C.~P.}\ \bibnamefont {Sun}},\
  }\bibfield  {title} {\emph {\bibinfo {title} {Photon blockade induced by
  atoms with {R}ydberg coupling},\ }}\href {\doibase
  10.1103/PhysRevA.87.023822} {\bibfield  {journal} {\bibinfo  {journal} {Phys.
  Rev. A}\ }\textbf {\bibinfo {volume} {87}},\ \bibinfo {pages} {023822}
  (\bibinfo {year} {2013})}\BibitemShut {NoStop}%
\bibitem [{\citenamefont {Liao}\ and\ \citenamefont
  {Law}(2010)}]{Liao2010correlated}%
  \BibitemOpen
  \bibfield  {author} {\bibinfo {author} {\bibfnamefont {J.-Q.}\ \bibnamefont
  {Liao}}\ and\ \bibinfo {author} {\bibfnamefont {C.~K.}\ \bibnamefont {Law}},\
  }\bibfield  {title} {\emph {\bibinfo {title} {Correlated two-photon transport
  in a one-dimensional waveguide side-coupled to a nonlinear cavity},\ }}\href
  {\doibase 10.1103/PhysRevA.82.053836} {\bibfield  {journal} {\bibinfo
  {journal} {Phys. Rev. A}\ }\textbf {\bibinfo {volume} {82}},\ \bibinfo
  {pages} {053836} (\bibinfo {year} {2010})}\BibitemShut {NoStop}%
\bibitem [{\citenamefont {Ghosh}\ and\ \citenamefont
  {Liew}(2019)}]{Ghosh2019Dynamical}%
  \BibitemOpen
  \bibfield  {author} {\bibinfo {author} {\bibfnamefont {S.}~\bibnamefont
  {Ghosh}}\ and\ \bibinfo {author} {\bibfnamefont {T.~C.~H.}\ \bibnamefont
  {Liew}},\ }\bibfield  {title} {\emph {\bibinfo {title} {Dynamical blockade in
  a single-mode bosonic system},\ }}\href {\doibase
  10.1103/PhysRevLett.123.013602} {\bibfield  {journal} {\bibinfo  {journal}
  {Phys. Rev. Lett.}\ }\textbf {\bibinfo {volume} {123}},\ \bibinfo {pages}
  {013602} (\bibinfo {year} {2019})}\BibitemShut {NoStop}%
\bibitem [{\citenamefont {L\"u}\ \emph {et~al.}(2015)\citenamefont {L\"u},
  \citenamefont {Wu}, \citenamefont {Johansson}, \citenamefont {Jing},
  \citenamefont {Zhang},\ and\ \citenamefont {Nori}}]{Lu2015Squeezed}%
  \BibitemOpen
  \bibfield  {author} {\bibinfo {author} {\bibfnamefont {X.-Y.}\ \bibnamefont
  {L\"u}}, \bibinfo {author} {\bibfnamefont {Y.}~\bibnamefont {Wu}}, \bibinfo
  {author} {\bibfnamefont {J.~R.}\ \bibnamefont {Johansson}}, \bibinfo {author}
  {\bibfnamefont {H.}~\bibnamefont {Jing}}, \bibinfo {author} {\bibfnamefont
  {J.}~\bibnamefont {Zhang}}, \ and\ \bibinfo {author} {\bibfnamefont
  {F.}~\bibnamefont {Nori}},\ }\bibfield  {title} {\emph {\bibinfo {title}
  {Squeezed optomechanics with phase-matched amplification and dissipation},\
  }}\href {\doibase 10.1103/PhysRevLett.114.093602} {\bibfield  {journal}
  {\bibinfo  {journal} {Phys. Rev. Lett.}\ }\textbf {\bibinfo {volume} {114}},\
  \bibinfo {pages} {093602} (\bibinfo {year} {2015})}\BibitemShut {NoStop}%
\bibitem [{\citenamefont {Peyronel}\ \emph {et~al.}(2012)\citenamefont
  {Peyronel}, \citenamefont {Firstenberg}, \citenamefont {Liang}, \citenamefont
  {Hofferberth}, \citenamefont {Gorshkov}, \citenamefont {Pohl}, \citenamefont
  {Lukin},\ and\ \citenamefont {Vuletić}}]{Peyronel2012quantum}%
  \BibitemOpen
  \bibfield  {author} {\bibinfo {author} {\bibfnamefont {T.}~\bibnamefont
  {Peyronel}}, \bibinfo {author} {\bibfnamefont {O.}~\bibnamefont
  {Firstenberg}}, \bibinfo {author} {\bibfnamefont {Q.-Y.}\ \bibnamefont
  {Liang}}, \bibinfo {author} {\bibfnamefont {S.}~\bibnamefont {Hofferberth}},
  \bibinfo {author} {\bibfnamefont {A.~V.}\ \bibnamefont {Gorshkov}}, \bibinfo
  {author} {\bibfnamefont {T.}~\bibnamefont {Pohl}}, \bibinfo {author}
  {\bibfnamefont {M.~D.}\ \bibnamefont {Lukin}}, \ and\ \bibinfo {author}
  {\bibfnamefont {V.}~\bibnamefont {Vuletić}},\ }\bibfield  {title} {\emph
  {\bibinfo {title} {Quantum nonlinear optics with single photons enabled by
  strongly interacting atoms},\ }}\href {\doibase 10.1038/nature113612}
  {\bibfield  {journal} {\bibinfo  {journal} {Nature (London)}\ }\textbf
  {\bibinfo {volume} {488}},\ \bibinfo {pages} {57} (\bibinfo {year}
  {2012})}\BibitemShut {NoStop}%
\bibitem [{\citenamefont {Reinhard}\ \emph {et~al.}(2012)\citenamefont
  {Reinhard}, \citenamefont {Volz}, \citenamefont {Winger}, \citenamefont
  {Badolato}, \citenamefont {Hennessy}, \citenamefont {Hu},\ and\ \citenamefont
  {Imamoğlu}}]{Reinhard2012strongly}%
  \BibitemOpen
  \bibfield  {author} {\bibinfo {author} {\bibfnamefont {A.}~\bibnamefont
  {Reinhard}}, \bibinfo {author} {\bibfnamefont {T.}~\bibnamefont {Volz}},
  \bibinfo {author} {\bibfnamefont {M.}~\bibnamefont {Winger}}, \bibinfo
  {author} {\bibfnamefont {A.}~\bibnamefont {Badolato}}, \bibinfo {author}
  {\bibfnamefont {K.~J.}\ \bibnamefont {Hennessy}}, \bibinfo {author}
  {\bibfnamefont {E.~L.}\ \bibnamefont {Hu}}, \ and\ \bibinfo {author}
  {\bibfnamefont {A.}~\bibnamefont {Imamoğlu}},\ }\bibfield  {title} {\emph
  {\bibinfo {title} {Strongly correlated photons on a chip},\ }}\href {\doibase
  10.1038/nphoton.2011.321} {\bibfield  {journal} {\bibinfo  {journal} {Nat.
  Photon.}\ }\textbf {\bibinfo {volume} {6}},\ \bibinfo {pages} {93} (\bibinfo
  {year} {2012})}\BibitemShut {NoStop}%
\bibitem [{\citenamefont {Liew}\ and\ \citenamefont
  {Savona}(2010)}]{Liew2010Single}%
  \BibitemOpen
  \bibfield  {author} {\bibinfo {author} {\bibfnamefont {T.~C.~H.}\
  \bibnamefont {Liew}}\ and\ \bibinfo {author} {\bibfnamefont {V.}~\bibnamefont
  {Savona}},\ }\bibfield  {title} {\emph {\bibinfo {title} {Single photons from
  coupled quantum modes},\ }}\href {\doibase 10.1103/PhysRevLett.104.183601}
  {\bibfield  {journal} {\bibinfo  {journal} {Phys. Rev. Lett.}\ }\textbf
  {\bibinfo {volume} {104}},\ \bibinfo {pages} {183601} (\bibinfo {year}
  {2010})}\BibitemShut {NoStop}%
\bibitem [{\citenamefont {Bamba}\ \emph {et~al.}(2011)\citenamefont {Bamba},
  \citenamefont {Imamo\ifmmode~\breve{g}\else \u{g}\fi{}lu}, \citenamefont
  {Carusotto},\ and\ \citenamefont {Ciuti}}]{Bamba2011Origin}%
  \BibitemOpen
  \bibfield  {author} {\bibinfo {author} {\bibfnamefont {M.}~\bibnamefont
  {Bamba}}, \bibinfo {author} {\bibfnamefont {A.}~\bibnamefont
  {Imamo\ifmmode~\breve{g}\else \u{g}\fi{}lu}}, \bibinfo {author}
  {\bibfnamefont {I.}~\bibnamefont {Carusotto}}, \ and\ \bibinfo {author}
  {\bibfnamefont {C.}~\bibnamefont {Ciuti}},\ }\bibfield  {title} {\emph
  {\bibinfo {title} {Origin of strong photon antibunching in weakly nonlinear
  photonic molecules},\ }}\href {\doibase 10.1103/PhysRevA.83.021802}
  {\bibfield  {journal} {\bibinfo  {journal} {Phys. Rev. A}\ }\textbf {\bibinfo
  {volume} {83}},\ \bibinfo {pages} {021802} (\bibinfo {year}
  {2011})}\BibitemShut {NoStop}%
\bibitem [{\citenamefont {Snijders}\ \emph {et~al.}(2018)\citenamefont
  {Snijders}, \citenamefont {Frey}, \citenamefont {Norman}, \citenamefont
  {Flayac}, \citenamefont {Savona}, \citenamefont {Gossard}, \citenamefont
  {Bowers}, \citenamefont {van Exter}, \citenamefont {Bouwmeester},\ and\
  \citenamefont {L\"offler}}]{Snijders2018Observation}%
  \BibitemOpen
  \bibfield  {author} {\bibinfo {author} {\bibfnamefont {H.~J.}\ \bibnamefont
  {Snijders}}, \bibinfo {author} {\bibfnamefont {J.~A.}\ \bibnamefont {Frey}},
  \bibinfo {author} {\bibfnamefont {J.}~\bibnamefont {Norman}}, \bibinfo
  {author} {\bibfnamefont {H.}~\bibnamefont {Flayac}}, \bibinfo {author}
  {\bibfnamefont {V.}~\bibnamefont {Savona}}, \bibinfo {author} {\bibfnamefont
  {A.~C.}\ \bibnamefont {Gossard}}, \bibinfo {author} {\bibfnamefont {J.~E.}\
  \bibnamefont {Bowers}}, \bibinfo {author} {\bibfnamefont {M.~P.}\
  \bibnamefont {van Exter}}, \bibinfo {author} {\bibfnamefont {D.}~\bibnamefont
  {Bouwmeester}}, \ and\ \bibinfo {author} {\bibfnamefont {W.}~\bibnamefont
  {L\"offler}},\ }\bibfield  {title} {\emph {\bibinfo {title} {Observation of
  the unconventional photon blockade},\ }}\href {\doibase
  10.1103/PhysRevLett.121.043601} {\bibfield  {journal} {\bibinfo  {journal}
  {Phys. Rev. Lett.}\ }\textbf {\bibinfo {volume} {121}},\ \bibinfo {pages}
  {043601} (\bibinfo {year} {2018})}\BibitemShut {NoStop}%
\bibitem [{\citenamefont {Vaneph}\ \emph {et~al.}(2018)\citenamefont {Vaneph},
  \citenamefont {Morvan}, \citenamefont {Aiello}, \citenamefont {F\'echant},
  \citenamefont {Aprili}, \citenamefont {Gabelli},\ and\ \citenamefont
  {Est\`eve}}]{Vaneph2018Observation}%
  \BibitemOpen
  \bibfield  {author} {\bibinfo {author} {\bibfnamefont {C.}~\bibnamefont
  {Vaneph}}, \bibinfo {author} {\bibfnamefont {A.}~\bibnamefont {Morvan}},
  \bibinfo {author} {\bibfnamefont {G.}~\bibnamefont {Aiello}}, \bibinfo
  {author} {\bibfnamefont {M.}~\bibnamefont {F\'echant}}, \bibinfo {author}
  {\bibfnamefont {M.}~\bibnamefont {Aprili}}, \bibinfo {author} {\bibfnamefont
  {J.}~\bibnamefont {Gabelli}}, \ and\ \bibinfo {author} {\bibfnamefont
  {J.}~\bibnamefont {Est\`eve}},\ }\bibfield  {title} {\emph {\bibinfo {title}
  {Observation of the unconventional photon blockade in the microwave domain},\
  }}\href {\doibase 10.1103/PhysRevLett.121.043602} {\bibfield  {journal}
  {\bibinfo  {journal} {Phys. Rev. Lett.}\ }\textbf {\bibinfo {volume} {121}},\
  \bibinfo {pages} {043602} (\bibinfo {year} {2018})}\BibitemShut {NoStop}%
\bibitem [{\citenamefont {Hartmann}\ and\ \citenamefont
  {Plenio}(2007)}]{Hartmann2007strong}%
  \BibitemOpen
  \bibfield  {author} {\bibinfo {author} {\bibfnamefont {M.~J.}\ \bibnamefont
  {Hartmann}}\ and\ \bibinfo {author} {\bibfnamefont {M.~B.}\ \bibnamefont
  {Plenio}},\ }\bibfield  {title} {\emph {\bibinfo {title} {Strong photon
  nonlinearities and photonic {M}ott insulators},\ }}\href {\doibase
  10.1103/PhysRevLett.99.103601} {\bibfield  {journal} {\bibinfo  {journal}
  {Phys. Rev. Lett.}\ }\textbf {\bibinfo {volume} {99}},\ \bibinfo {pages}
  {103601} (\bibinfo {year} {2007})}\BibitemShut {NoStop}%
\bibitem [{\citenamefont {Chang}\ \emph {et~al.}(2014)\citenamefont {Chang},
  \citenamefont {Vuletić},\ and\ \citenamefont {Lukin}}]{Chang2014Quantum}%
  \BibitemOpen
  \bibfield  {author} {\bibinfo {author} {\bibfnamefont {D.~E.}\ \bibnamefont
  {Chang}}, \bibinfo {author} {\bibfnamefont {V.}~\bibnamefont {Vuletić}}, \
  and\ \bibinfo {author} {\bibfnamefont {M.~D.}\ \bibnamefont {Lukin}},\
  }\bibfield  {title} {\emph {\bibinfo {title} {Quantum nonlinear
  optics—photon by photon},\ }}\href {\doibase 10.1038/nphoton.2014.192}
  {\bibfield  {journal} {\bibinfo  {journal} {Nat. Photon.}\ }\textbf {\bibinfo
  {volume} {8}},\ \bibinfo {pages} {685} (\bibinfo {year} {2014})}\BibitemShut
  {NoStop}%
\bibitem [{\citenamefont {Imamo\ifmmode~\bar{g}\else \={g}\fi{}lu}\ \emph
  {et~al.}(1997)\citenamefont {Imamo\ifmmode~\bar{g}\else \={g}\fi{}lu},
  \citenamefont {Schmidt}, \citenamefont {Woods},\ and\ \citenamefont
  {Deutsch}}]{Imamo1997strongly}%
  \BibitemOpen
  \bibfield  {author} {\bibinfo {author} {\bibfnamefont {A.}~\bibnamefont
  {Imamo\ifmmode~\bar{g}\else \={g}\fi{}lu}}, \bibinfo {author} {\bibfnamefont
  {H.}~\bibnamefont {Schmidt}}, \bibinfo {author} {\bibfnamefont
  {G.}~\bibnamefont {Woods}}, \ and\ \bibinfo {author} {\bibfnamefont
  {M.}~\bibnamefont {Deutsch}},\ }\bibfield  {title} {\emph {\bibinfo {title}
  {Strongly interacting photons in a nonlinear cavity},\ }}\href {\doibase
  10.1103/PhysRevLett.79.1467} {\bibfield  {journal} {\bibinfo  {journal}
  {Phys. Rev. Lett.}\ }\textbf {\bibinfo {volume} {79}},\ \bibinfo {pages}
  {1467} (\bibinfo {year} {1997})}\BibitemShut {NoStop}%
\bibitem [{\citenamefont {Tang}\ \emph {et~al.}(2015)\citenamefont {Tang},
  \citenamefont {Geng},\ and\ \citenamefont {Xu}}]{Tang2015Quantum}%
  \BibitemOpen
  \bibfield  {author} {\bibinfo {author} {\bibfnamefont {J.}~\bibnamefont
  {Tang}}, \bibinfo {author} {\bibfnamefont {W.}~\bibnamefont {Geng}}, \ and\
  \bibinfo {author} {\bibfnamefont {X.}~\bibnamefont {Xu}},\ }\bibfield
  {title} {\emph {\bibinfo {title} {Quantum interference induced photon
  blockade in a coupled single quantum dot-cavity system},\ }}\href {\doibase
  10.1038/srep09252} {\bibfield  {journal} {\bibinfo  {journal} {Sci. Rep.}\
  }\textbf {\bibinfo {volume} {5}},\ \bibinfo {pages} {9252} (\bibinfo {year}
  {2015})}\BibitemShut {NoStop}%
\bibitem [{\citenamefont {Carmichael}(1999)}]{Carmichael1999statistical}%
  \BibitemOpen
  \bibfield  {author} {\bibinfo {author} {\bibfnamefont {H.}~\bibnamefont
  {Carmichael}},\ }\href@noop {} {\emph {\bibinfo {title} {Statistical Methods
  in Quantum Optics}}}\ (\bibinfo  {publisher} {Springer, Berlin},\ \bibinfo
  {year} {1999})\BibitemShut {NoStop}%
\bibitem [{\citenamefont {Wang}\ \emph {et~al.}(2020)\citenamefont {Wang},
  \citenamefont {Bai}, \citenamefont {Liu}, \citenamefont {Zhang},\ and\
  \citenamefont {Wang}}]{Wang2020Photon}%
  \BibitemOpen
  \bibfield  {author} {\bibinfo {author} {\bibfnamefont {D.-Y.}\ \bibnamefont
  {Wang}}, \bibinfo {author} {\bibfnamefont {C.-H.}\ \bibnamefont {Bai}},
  \bibinfo {author} {\bibfnamefont {S.}~\bibnamefont {Liu}}, \bibinfo {author}
  {\bibfnamefont {S.}~\bibnamefont {Zhang}}, \ and\ \bibinfo {author}
  {\bibfnamefont {H.-F.}\ \bibnamefont {Wang}},\ }\bibfield  {title} {\emph
  {\bibinfo {title} {Photon blockade in a double-cavity optomechanical system
  with nonreciprocal coupling},\ }}\href {\doibase 10.1088/1367-2630/abaa8a}
  {\bibfield  {journal} {\bibinfo  {journal} {New J. Phys.}\ }\textbf {\bibinfo
  {volume} {22}},\ \bibinfo {pages} {093006} (\bibinfo {year}
  {2020})}\BibitemShut {NoStop}%
\bibitem [{\citenamefont {Jaynes}\ and\ \citenamefont
  {Cummings}(1963)}]{Jaynes1963comparison}%
  \BibitemOpen
  \bibfield  {author} {\bibinfo {author} {\bibfnamefont {E.}~\bibnamefont
  {Jaynes}}\ and\ \bibinfo {author} {\bibfnamefont {F.}~\bibnamefont
  {Cummings}},\ }\bibfield  {title} {\emph {\bibinfo {title} {Comparison of
  quantum and semiclassical radiation theories with application to the beam
  maser},\ }}\href {\doibase 10.1109/PROC.1963.1664} {\bibfield  {journal}
  {\bibinfo  {journal} {Proc. IEEE}\ }\textbf {\bibinfo {volume} {51}},\
  \bibinfo {pages} {89} (\bibinfo {year} {1963})}\BibitemShut {NoStop}%
\bibitem [{\citenamefont {Shevchuk}\ \emph {et~al.}(2017)\citenamefont
  {Shevchuk}, \citenamefont {Steele},\ and\ \citenamefont
  {Blanter}}]{Shevchuk2017Strong}%
  \BibitemOpen
  \bibfield  {author} {\bibinfo {author} {\bibfnamefont {O.}~\bibnamefont
  {Shevchuk}}, \bibinfo {author} {\bibfnamefont {G.~A.}\ \bibnamefont
  {Steele}}, \ and\ \bibinfo {author} {\bibfnamefont {Y.~M.}\ \bibnamefont
  {Blanter}},\ }\bibfield  {title} {\emph {\bibinfo {title} {Strong and tunable
  couplings in flux-mediated optomechanics},\ }}\href {\doibase
  10.1103/PhysRevB.96.014508} {\bibfield  {journal} {\bibinfo  {journal} {Phys.
  Rev. B}\ }\textbf {\bibinfo {volume} {96}},\ \bibinfo {pages} {014508}
  (\bibinfo {year} {2017})}\BibitemShut {NoStop}%
\bibitem [{\citenamefont {Rodrigues}\ \emph {et~al.}(2019)\citenamefont
  {Rodrigues}, \citenamefont {Bothner},\ and\ \citenamefont
  {Steele}}]{Rodrigues2019coupling}%
  \BibitemOpen
  \bibfield  {author} {\bibinfo {author} {\bibfnamefont {I.~C.}\ \bibnamefont
  {Rodrigues}}, \bibinfo {author} {\bibfnamefont {D.}~\bibnamefont {Bothner}},
  \ and\ \bibinfo {author} {\bibfnamefont {G.~A.}\ \bibnamefont {Steele}},\
  }\bibfield  {title} {\emph {\bibinfo {title} {Coupling microwave photons to a
  mechanical resonator using quantum interference},\ }}\href {\doibase
  10.1038/s41467-019-12964-2} {\bibfield  {journal} {\bibinfo  {journal} {Nat.
  Commun.}\ }\textbf {\bibinfo {volume} {10}},\ \bibinfo {pages} {5359}
  (\bibinfo {year} {2019})}\BibitemShut {NoStop}%
\bibitem [{\citenamefont {Kounalakis}\ \emph {et~al.}(2020)\citenamefont
  {Kounalakis}, \citenamefont {Blanter},\ and\ \citenamefont
  {Steele}}]{Kounalakis2020Flux}%
  \BibitemOpen
  \bibfield  {author} {\bibinfo {author} {\bibfnamefont {M.}~\bibnamefont
  {Kounalakis}}, \bibinfo {author} {\bibfnamefont {Y.~M.}\ \bibnamefont
  {Blanter}}, \ and\ \bibinfo {author} {\bibfnamefont {G.~A.}\ \bibnamefont
  {Steele}},\ }\bibfield  {title} {\emph {\bibinfo {title} {Flux-mediated
  optomechanics with a transmon qubit in the single-photon ultrastrong-coupling
  regime},\ }}\href {\doibase 10.1103/PhysRevResearch.2.023335} {\bibfield
  {journal} {\bibinfo  {journal} {Phys. Rev. Res.}\ }\textbf {\bibinfo {volume}
  {2}},\ \bibinfo {pages} {023335} (\bibinfo {year} {2020})}\BibitemShut
  {NoStop}%
\bibitem [{\citenamefont {Wang}\ \emph {et~al.}(2019)\citenamefont {Wang},
  \citenamefont {Zhang}, \citenamefont {Xu}, \citenamefont {Li}, \citenamefont
  {Zhu}, \citenamefont {Tsai},\ and\ \citenamefont {You}}]{wang2019simulation}%
  \BibitemOpen
  \bibfield  {author} {\bibinfo {author} {\bibfnamefont {Y.-P.}\ \bibnamefont
  {Wang}}, \bibinfo {author} {\bibfnamefont {G.-Q.}\ \bibnamefont {Zhang}},
  \bibinfo {author} {\bibfnamefont {D.}~\bibnamefont {Xu}}, \bibinfo {author}
  {\bibfnamefont {T.-F.}\ \bibnamefont {Li}}, \bibinfo {author} {\bibfnamefont
  {S.-Y.}\ \bibnamefont {Zhu}}, \bibinfo {author} {\bibfnamefont {J.~S.}\
  \bibnamefont {Tsai}}, \ and\ \bibinfo {author} {\bibfnamefont {J.~Q.}\
  \bibnamefont {You}},\ }\bibfield  {title} {\emph {\bibinfo {title} {Quantum
  simulation of the fermion-boson composite quasi-particles with a driven
  qubit-magnon hybrid quantum system},\ }}\href {\doibase
  10.48550/ARXIV.1903.12498} {\bibfield  {journal} {\bibinfo  {journal}
  {arXiv}\ }\textbf {\bibinfo {volume} {1903}},\ \bibinfo {pages} {12498}
  (\bibinfo {year} {2019})}\BibitemShut {NoStop}%
\bibitem [{\citenamefont {Wang}\ \emph {et~al.}(2016)\citenamefont {Wang},
  \citenamefont {Zhang}, \citenamefont {Zhang}, \citenamefont {Luo},
  \citenamefont {Xiong}, \citenamefont {Wang}, \citenamefont {Li},
  \citenamefont {Hu},\ and\ \citenamefont {You}}]{wang2016magnon}%
  \BibitemOpen
  \bibfield  {author} {\bibinfo {author} {\bibfnamefont {Y.-P.}\ \bibnamefont
  {Wang}}, \bibinfo {author} {\bibfnamefont {G.-Q.}\ \bibnamefont {Zhang}},
  \bibinfo {author} {\bibfnamefont {D.}~\bibnamefont {Zhang}}, \bibinfo
  {author} {\bibfnamefont {X.-Q.}\ \bibnamefont {Luo}}, \bibinfo {author}
  {\bibfnamefont {W.}~\bibnamefont {Xiong}}, \bibinfo {author} {\bibfnamefont
  {S.-P.}\ \bibnamefont {Wang}}, \bibinfo {author} {\bibfnamefont {T.-F.}\
  \bibnamefont {Li}}, \bibinfo {author} {\bibfnamefont {C.-M.}\ \bibnamefont
  {Hu}}, \ and\ \bibinfo {author} {\bibfnamefont {J.~Q.}\ \bibnamefont {You}},\
  }\bibfield  {title} {\emph {\bibinfo {title} {Magnon {K}err effect in a
  strongly coupled cavity-magnon system},\ }}\href {\doibase
  10.1103/PhysRevB.94.224410} {\bibfield  {journal} {\bibinfo  {journal} {Phys.
  Rev. B}\ }\textbf {\bibinfo {volume} {94}},\ \bibinfo {pages} {224410}
  (\bibinfo {year} {2016})}\BibitemShut {NoStop}%
\bibitem [{\citenamefont {Scully}\ and\ \citenamefont
  {Zubairy}(1997)}]{Scully1997quantum}%
  \BibitemOpen
  \bibfield  {author} {\bibinfo {author} {\bibfnamefont {M.~O.}\ \bibnamefont
  {Scully}}\ and\ \bibinfo {author} {\bibfnamefont {M.~S.}\ \bibnamefont
  {Zubairy}},\ }\href@noop {} {\emph {\bibinfo {title} {Quantum Optics}}}\
  (\bibinfo  {publisher} {Cambridge University Press, Cambridge},\ \bibinfo
  {year} {1997})\BibitemShut {NoStop}%
\bibitem [{\citenamefont {Eleuch}(2008)}]{Eleuch2008Photon}%
  \BibitemOpen
  \bibfield  {author} {\bibinfo {author} {\bibfnamefont {H.}~\bibnamefont
  {Eleuch}},\ }\bibfield  {title} {\emph {\bibinfo {title} {Photon statistics
  of light in semiconductor microcavities},\ }}\href {\doibase
  10.1088/0953-4075/41/5/055502} {\bibfield  {journal} {\bibinfo  {journal} {J.
  Phys. B}\ }\textbf {\bibinfo {volume} {41}},\ \bibinfo {pages} {055502}
  (\bibinfo {year} {2008})}\BibitemShut {NoStop}%
\bibitem [{\citenamefont {Rabl}(2011)}]{Rabl2011photon}%
  \BibitemOpen
  \bibfield  {author} {\bibinfo {author} {\bibfnamefont {P.}~\bibnamefont
  {Rabl}},\ }\bibfield  {title} {\emph {\bibinfo {title} {Photon blockade
  effect in optomechanical systems},\ }}\href {\doibase
  10.1103/PhysRevLett.107.063601} {\bibfield  {journal} {\bibinfo  {journal}
  {Phys. Rev. Lett.}\ }\textbf {\bibinfo {volume} {107}},\ \bibinfo {pages}
  {063601} (\bibinfo {year} {2011})}\BibitemShut {NoStop}%
\bibitem [{\citenamefont {K\'om\'ar}\ \emph {et~al.}(2013)\citenamefont
  {K\'om\'ar}, \citenamefont {Bennett}, \citenamefont {Stannigel},
  \citenamefont {Habraken}, \citenamefont {Rabl}, \citenamefont {Zoller},\ and\
  \citenamefont {Lukin}}]{Kar2013Single}%
  \BibitemOpen
  \bibfield  {author} {\bibinfo {author} {\bibfnamefont {P.}~\bibnamefont
  {K\'om\'ar}}, \bibinfo {author} {\bibfnamefont {S.~D.}\ \bibnamefont
  {Bennett}}, \bibinfo {author} {\bibfnamefont {K.}~\bibnamefont {Stannigel}},
  \bibinfo {author} {\bibfnamefont {S.~J.~M.}\ \bibnamefont {Habraken}},
  \bibinfo {author} {\bibfnamefont {P.}~\bibnamefont {Rabl}}, \bibinfo {author}
  {\bibfnamefont {P.}~\bibnamefont {Zoller}}, \ and\ \bibinfo {author}
  {\bibfnamefont {M.~D.}\ \bibnamefont {Lukin}},\ }\bibfield  {title} {\emph
  {\bibinfo {title} {Single-photon nonlinearities in two-mode optomechanics},\
  }}\href {\doibase 10.1103/PhysRevA.87.013839} {\bibfield  {journal} {\bibinfo
   {journal} {Phys. Rev. A}\ }\textbf {\bibinfo {volume} {87}},\ \bibinfo
  {pages} {013839} (\bibinfo {year} {2013})}\BibitemShut {NoStop}%
\bibitem [{\citenamefont {Tang}\ \emph {et~al.}(2019)\citenamefont {Tang},
  \citenamefont {Deng},\ and\ \citenamefont {Lee}}]{Tang2019Strong}%
  \BibitemOpen
  \bibfield  {author} {\bibinfo {author} {\bibfnamefont {J.}~\bibnamefont
  {Tang}}, \bibinfo {author} {\bibfnamefont {Y.}~\bibnamefont {Deng}}, \ and\
  \bibinfo {author} {\bibfnamefont {C.}~\bibnamefont {Lee}},\ }\bibfield
  {title} {\emph {\bibinfo {title} {Strong photon blockade mediated by optical
  stark shift in a single-atom--cavity system},\ }}\href {\doibase
  10.1103/PhysRevApplied.12.044065} {\bibfield  {journal} {\bibinfo  {journal}
  {Phys. Rev. Applied}\ }\textbf {\bibinfo {volume} {12}},\ \bibinfo {pages}
  {044065} (\bibinfo {year} {2019})}\BibitemShut {NoStop}%
\bibitem [{\citenamefont {Li}\ \emph {et~al.}(2021)\citenamefont {Li},
  \citenamefont {Li},\ and\ \citenamefont {Zhong}}]{Li2021strong}%
  \BibitemOpen
  \bibfield  {author} {\bibinfo {author} {\bibfnamefont {Z.}~\bibnamefont
  {Li}}, \bibinfo {author} {\bibfnamefont {X.}~\bibnamefont {Li}}, \ and\
  \bibinfo {author} {\bibfnamefont {X.}~\bibnamefont {Zhong}},\ }\bibfield
  {title} {\emph {\bibinfo {title} {Strong photon blockade in an all-fiber
  emitter-cavity quantum electrodynamics system},\ }}\href {\doibase
  10.1103/PhysRevA.103.043724} {\bibfield  {journal} {\bibinfo  {journal}
  {Phys. Rev. A}\ }\textbf {\bibinfo {volume} {103}},\ \bibinfo {pages}
  {043724} (\bibinfo {year} {2021})}\BibitemShut {NoStop}%
\bibitem [{\citenamefont {Schuster}\ \emph {et~al.}(2007)\citenamefont
  {Schuster}, \citenamefont {Houck}, \citenamefont {Schreier}, \citenamefont
  {Wallraff}, \citenamefont {Gambetta}, \citenamefont {Blais}, \citenamefont
  {Frunzio}, \citenamefont {Majer}, \citenamefont {Johnson}, \citenamefont
  {Devoret}, \citenamefont {Girvin},\ and\ \citenamefont
  {Schoelkopf}}]{Schuster2007Resolving}%
  \BibitemOpen
  \bibfield  {author} {\bibinfo {author} {\bibfnamefont {D.~I.}\ \bibnamefont
  {Schuster}}, \bibinfo {author} {\bibfnamefont {A.~A.}\ \bibnamefont {Houck}},
  \bibinfo {author} {\bibfnamefont {J.~A.}\ \bibnamefont {Schreier}}, \bibinfo
  {author} {\bibfnamefont {A.}~\bibnamefont {Wallraff}}, \bibinfo {author}
  {\bibfnamefont {J.~M.}\ \bibnamefont {Gambetta}}, \bibinfo {author}
  {\bibfnamefont {A.}~\bibnamefont {Blais}}, \bibinfo {author} {\bibfnamefont
  {L.}~\bibnamefont {Frunzio}}, \bibinfo {author} {\bibfnamefont
  {J.}~\bibnamefont {Majer}}, \bibinfo {author} {\bibfnamefont
  {B.}~\bibnamefont {Johnson}}, \bibinfo {author} {\bibfnamefont {M.~H.}\
  \bibnamefont {Devoret}}, \bibinfo {author} {\bibfnamefont {S.~M.}\
  \bibnamefont {Girvin}}, \ and\ \bibinfo {author} {\bibfnamefont {R.~J.}\
  \bibnamefont {Schoelkopf}},\ }\bibfield  {title} {\emph {\bibinfo {title}
  {Resolving photon number states in a superconducting circuit},\ }}\href
  {\doibase 10.1038/nature05461} {\bibfield  {journal} {\bibinfo  {journal}
  {Nature (London)}\ }\textbf {\bibinfo {volume} {445}},\ \bibinfo {pages}
  {515} (\bibinfo {year} {2007})}\BibitemShut {NoStop}%
\bibitem [{\citenamefont {Vlastakis}\ \emph {et~al.}(2013)\citenamefont
  {Vlastakis}, \citenamefont {Kirchmair}, \citenamefont {Leghtas},
  \citenamefont {E.~Nigg}, \citenamefont {Frunzio}, \citenamefont {Girvin},
  \citenamefont {Mirrahimi}, \citenamefont {Devoret},\ and\ \citenamefont
  {Schoelkopf}}]{Vlastakis2013Deterministically}%
  \BibitemOpen
  \bibfield  {author} {\bibinfo {author} {\bibfnamefont {B.}~\bibnamefont
  {Vlastakis}}, \bibinfo {author} {\bibfnamefont {G.}~\bibnamefont
  {Kirchmair}}, \bibinfo {author} {\bibfnamefont {Z.}~\bibnamefont {Leghtas}},
  \bibinfo {author} {\bibfnamefont {S.}~\bibnamefont {E.~Nigg}}, \bibinfo
  {author} {\bibfnamefont {L.}~\bibnamefont {Frunzio}}, \bibinfo {author}
  {\bibfnamefont {S.~M.}\ \bibnamefont {Girvin}}, \bibinfo {author}
  {\bibfnamefont {M.}~\bibnamefont {Mirrahimi}}, \bibinfo {author}
  {\bibfnamefont {M.~H.}\ \bibnamefont {Devoret}}, \ and\ \bibinfo {author}
  {\bibfnamefont {R.~J.}\ \bibnamefont {Schoelkopf}},\ }\bibfield  {title}
  {\emph {\bibinfo {title} {Deterministically encoding quantum information
  using 100-photon {S}chrödinger cat states},\ }}\href {\doibase
  10.1126/science.1243289} {\bibfield  {journal} {\bibinfo  {journal}
  {Science}\ }\textbf {\bibinfo {volume} {342}},\ \bibinfo {pages} {607}
  (\bibinfo {year} {2013})}\BibitemShut {NoStop}%
\bibitem [{\citenamefont {Vlastakis}\ \emph {et~al.}(2015)\citenamefont
  {Vlastakis}, \citenamefont {Petrenko}, \citenamefont {Ofek}, \citenamefont
  {Sun}, \citenamefont {Leghtas}, \citenamefont {Sliwa}, \citenamefont {Liu},
  \citenamefont {Hatridge}, \citenamefont {Blumoff}, \citenamefont {Frunzio},
  \citenamefont {Mirrahimi}, \citenamefont {Jiang}, \citenamefont {Devoret},\
  and\ \citenamefont {Schoelkopf}}]{Vlastakis2015Characterizing}%
  \BibitemOpen
  \bibfield  {author} {\bibinfo {author} {\bibfnamefont {B.}~\bibnamefont
  {Vlastakis}}, \bibinfo {author} {\bibfnamefont {A.}~\bibnamefont {Petrenko}},
  \bibinfo {author} {\bibfnamefont {N.}~\bibnamefont {Ofek}}, \bibinfo {author}
  {\bibfnamefont {L.}~\bibnamefont {Sun}}, \bibinfo {author} {\bibfnamefont
  {Z.}~\bibnamefont {Leghtas}}, \bibinfo {author} {\bibfnamefont
  {K.}~\bibnamefont {Sliwa}}, \bibinfo {author} {\bibfnamefont
  {Y.}~\bibnamefont {Liu}}, \bibinfo {author} {\bibfnamefont {M.}~\bibnamefont
  {Hatridge}}, \bibinfo {author} {\bibfnamefont {J.}~\bibnamefont {Blumoff}},
  \bibinfo {author} {\bibfnamefont {L.}~\bibnamefont {Frunzio}}, \bibinfo
  {author} {\bibfnamefont {M.}~\bibnamefont {Mirrahimi}}, \bibinfo {author}
  {\bibfnamefont {L.}~\bibnamefont {Jiang}}, \bibinfo {author} {\bibfnamefont
  {M.~H.}\ \bibnamefont {Devoret}}, \ and\ \bibinfo {author} {\bibfnamefont
  {R.~J.}\ \bibnamefont {Schoelkopf}},\ }\bibfield  {title} {\emph {\bibinfo
  {title} {Characterizing entanglement of an artificial atom and a cavity cat
  state with {B}ell’s inequality},\ }}\href {\doibase 10.1038/ncomms9970}
  {\bibfield  {journal} {\bibinfo  {journal} {Nat. Commun.}\ }\textbf {\bibinfo
  {volume} {6}},\ \bibinfo {pages} {8970} (\bibinfo {year} {2015})}\BibitemShut
  {NoStop}%
\bibitem [{\citenamefont {Langford}\ \emph {et~al.}(2017)\citenamefont
  {Langford}, \citenamefont {Sagastizabal}, \citenamefont {Kounalakis},
  \citenamefont {Dickel}, \citenamefont {Bruno}, \citenamefont {Luthi},
  \citenamefont {Thoen}, \citenamefont {Endo},\ and\ \citenamefont
  {DiCarlo}}]{Langford2017Experimentally}%
  \BibitemOpen
  \bibfield  {author} {\bibinfo {author} {\bibfnamefont {N.~K.}\ \bibnamefont
  {Langford}}, \bibinfo {author} {\bibfnamefont {R.}~\bibnamefont
  {Sagastizabal}}, \bibinfo {author} {\bibfnamefont {M.}~\bibnamefont
  {Kounalakis}}, \bibinfo {author} {\bibfnamefont {C.}~\bibnamefont {Dickel}},
  \bibinfo {author} {\bibfnamefont {A.}~\bibnamefont {Bruno}}, \bibinfo
  {author} {\bibfnamefont {F.}~\bibnamefont {Luthi}}, \bibinfo {author}
  {\bibfnamefont {D.~J.}\ \bibnamefont {Thoen}}, \bibinfo {author}
  {\bibfnamefont {A.}~\bibnamefont {Endo}}, \ and\ \bibinfo {author}
  {\bibfnamefont {L.}~\bibnamefont {DiCarlo}},\ }\bibfield  {title} {\emph
  {\bibinfo {title} {Experimentally simulating the dynamics of quantum light
  and matter at deep-strong coupling},\ }}\href {\doibase
  10.1038/s41467-017-01061-x} {\bibfield  {journal} {\bibinfo  {journal} {Nat.
  Commun.}\ }\textbf {\bibinfo {volume} {8}},\ \bibinfo {pages} {1715}
  (\bibinfo {year} {2017})}\BibitemShut {NoStop}%
\bibitem [{\citenamefont {Edwards}(1984)}]{Harold1984Galois}%
  \BibitemOpen
  \bibfield  {author} {\bibinfo {author} {\bibfnamefont {H.~M.}\ \bibnamefont
  {Edwards}},\ }\href@noop {} {\emph {\bibinfo {title} {Galois Theory}}}\
  (\bibinfo  {publisher} {Springer, New York},\ \bibinfo {year}
  {1984})\BibitemShut {NoStop}%
\end{thebibliography}%
\end{document}